\documentclass[aos,preprint]{imsart}
\RequirePackage[OT1]{fontenc}
\RequirePackage{amsthm,amsmath}
\RequirePackage[numbers]{natbib}
\RequirePackage[colorlinks,citecolor=blue,urlcolor=blue]{hyperref}
\usepackage{amssymb,makeidx,latexsym,rotate,epsf,amssymb,graphicx,color, subfigure, url}

\newtheorem{ex}{Example}

\newtheorem{prop}{Proposition}

\newtheorem{df}{Definition}
\newtheorem{rmk}{Remark}

\arxiv{arXiv:0000.0000}

\startlocaldefs
\numberwithin{equation}{section}
\theoremstyle{plain}
\newtheorem{thm}{Theorem}[section]
\endlocaldefs

\begin{document}

\begin{frontmatter}
\title{Flexible extreme value inference and  Hill plots for small, mid and large samples}
\runtitle{Flexible extreme value inference and  Hill plots}
\thankstext{T1}{We acknowledge  support by Fondecyt Proyecto Regular N 1151441.}

\begin{aug}
\author{\fnms{Pavlina} \snm{Jordanova}\thanksref{m1}\ead[label = e2]{E-Mail: pavlina\_kj@abv.bg}}
\and
\author{\fnms{Milan} \snm{Stehl\'\i k}\thanksref{m2}\ead[label=e1]{E-Mail: mlnstehlik@gmail.com}},


\runauthor{P. Jordanova and M. Stehl\'\i k}

\affiliation{Shumen University\thanksmark{m1} and Universidad T\'ecnica Federico Santa Mar\'{\i}a\thanksmark{m2}}

\address{Departamento de  Matem\'atica \\ Universidad T\'ecnica Federico Santa Mar\'{\i}a\\ Casilla V 110, Valpara\'iso,  Chile \\
Department of Applied Statistics \\Johannes Kepler University Linz\\ Altenbergerstrasse 69, 4040 Linz, Austria\\
\printead*{e1}}

\address{Faculty  of Mathematics and Informatics\\ Shumen University\\ Bulgaria\\
\printead*{e2}}
\end{aug}

\begin{abstract}

Asymptotic normality of extreme value tail estimators  received much attention in the literature, giving rise to increasingly
complicated 2nd order regularity conditions. However, such conditions are really difficult to be checked for real data.
Especially it is difficult or impossible to  check  such conditions  using small samples.
Beside that most of those conditions suffer from the drawback of a potentially singular integral representations.
However, we can have various orders of approximation by normal distributions, e.g. Berry-Esseen Types and Edgeworth types.
In this paper we indicate that for Berry-Esseen Types of normal approximation and related asymptotic normality of generalized Hill estimators,
we do not necessarily need 2nd order regularity conditions and we can apply only Karamata's representation for regularly varying tails.
2nd order regularity conditions  however better relates to Edgeworth types of normal approximations, albeit requiring larger data samples for their
proper check. Finally both expansions are prone for bootstrap and other subsampling techniques.
All existing results indicate that proper
representation of tail behavior  play a special and somewhat intriguing role in that context.
We dispel that widespread opinion by providing a full characterization and representation, in a general regular variation
context, of the integral singularity phenomenon, highlighting its relation to an asymptotical normality of the Generalized Hill estimator without the 2nd order condition.
Thus application of this new methodology is  simple and much more flexible, optimal for real data sets.
Alternative and powerful versions of the Hill plot are also introduced and illustrated on ecological  data of snow extremes from Slovakia.

\end{abstract}


\end{frontmatter}

\section{Introduction and preliminaries}

Statistical Models for extreme value distributions have become increasingly popular in recent years, as they
provide a much better fit for data presenting some departures from normality.
Statistical inference for extreme value distribution typically requires a 2nd order regularity conditions, following works of  [de Haan and Stadtm\"uller (1996)],  [de Haan and  Ferreira (2006)
]  and [Geluk et al. (1997)].
However, 
\newpage
to check 2nd order regularity conditions is difficult for a real data, despite effort of some recent papers,  and this is also probably one of the main reasons why extreme value theory became highly complicated.
As we illustrate in this paper, this complexity can be reduced significantly.
Thus we suggest a more flexible approach to estimate extreme value index,  which is also illustrated both on real and synthetic data. The newly introduced methodology is based on Karamata's
representation. Several works on representations have been published
(\cite{Hall1978}, \cite{Hall82},\cite{Lo}, among others).
We acknowledge also the developing of 2nd order condition methods, since we learned much  from this methodology and, e.g. many comparisons based on such methods have appeared. {In  particular, Edgeworth expansion for the Hill estimator has been developed under 2nd order RV framework,
see e.g. \cite{ChengPan1998}.} However, 2nd order condition does not necessary hold (see e.g. \cite{Lo}) and therefore we work without this assumption.

{} We denote by $X_1, X_2, ..., X_n$ independent identically distributed (i.i.d) random variables (r.vs) with cumulative distribution function (c.d.f.) $F$ such that $\bar F \in  RV_{-\alpha}$ with  $\alpha > 0$. The last means that
there exist a positive limit $$\lim_{t\to \infty}\frac{\bar F(tx)}{\bar F(t)}$$  for all $x > 0.$
It is known that it is equivalent to $$\lim_{t\to \infty}\frac{\bar F(tx)}{\bar F(t)} = x^{-\alpha}$$ for $x>0$ with some $\alpha > 0$. The number $-\alpha$ is called the index of regular variation.

Denote the corresponding increasing order
statistics by $$X_{(1,n)} \leq X_{(2,n)} \leq ... \leq X_{(n,n)}{}$$
{ and by
\begin{equation}\label{HXknp}
H_{X, k, n, p} : = \frac{1}{k}\sum_{i=1}^{k}\left(\frac{X_{(n - i + 1, n)}}{X_{(n - k,
n)}}\right)^p  \quad p \in \mathbb{R}, \quad p \not= 0.
\end{equation}

The generalized Hill estimator (see \cite{BeranSSt} for original version with different and constrained parametrization) is defined by
$$\widehat{\gamma}_{X, k, n, p} : = \frac{1}{p}\left(1- \left[\frac{1}{k}\sum_{i=1}^{k}\left(\frac{X_{(n - i + 1, n)}}{X_{(n - k,
n)}}\right)^p \right]^{-1} \right) = \frac{1}{p}\left(1- H_{X, k, n, p}^{-1} \right)  \quad p \in \mathbb{R}, \quad p \not= 0.$$}

For $p = 0$ we consider limit, which is { the} well-known Hill estimator (see \cite{Hill}){,} defined as
{ $$\widehat{\gamma}_{X, k, n, 0} : =  \frac{1}{k}\sum_{i=1}^{k}\ln \left(\frac{X_{(n - i + 1, n)}}{X_{(n - k,
n)}}\right).$$}

{ In this paper we determine the exact distribution of $H_{X, k, n, p}$ for all $n \in \mathbb{N}$, $k = 1, 2, ..., n$ and Pareto distributed random variables. For any fixed $k = 1, 2, ..., $ and $n \to \infty$ we find appropriate normalizations, with non-random centering, such  that transformed $H_{X, k, n, p}$ and $\widehat{\gamma}_{X, k, n, p}$ are asymptotically standard normal. More generally we show that for the case when the distribution of the observed random variable has regularly varying tail and, without using the second order regularly varying condition, the limiting distribution for $n \to \infty$ and then $k \to \infty$ is again standard normal.}

The paper is organized as follows.  First we recall important definitions from Extreme Value Theory and consider the relation between the classes of distributions that achieve asymptotic normality of the normalized Hill estimator and the second order regularly varying condition. We present several examples that show that these classes of distributions are not equivalent. In  section 2 we consider the behavior of the Generalized Hill estimator, in case when the number of the order statistics is fixed and show that in Pareto case the distribution of $H_{X, k, n, p}$ coincides with the average of specific powers of uniformly distributed random variables. For $p = -\alpha$ this distribution is Irwin-Hall distribution. In comparison with the Hill estimator  where this distribution is Gamma with parameters $k$ and $k\alpha$. In more general case, when the tail of the distribution of the observed random variable is regularly varying these distribution appear in corresponding limits, when the sample size increases unboundedly. In the { end} of the second section we show by simulations that also for one of the most difficult cases for estimation, when the distribution of the observed random variable has very slowly regularly varying tail, and more precisely when it is Hill horror distributed, the considerations for fixed sample of order statistics and their plot for increasing $n$ are more informative than the Hill plots. The third section considers the cases when we obtain asymptotic normality of the Generalized Hill estimator and prove that in Pareto case we could achieve it also for the number of order statistics that is close to the sample size. In Theorem 3 we prove that the corresponding results about the Hill estimator could be considered as a particular case of the Generalized Hill estimator, therefore we call $\widehat{\gamma}_{X, k, n, p}$ in this way. In Section 4 we try to find "the most appropriate" case of $p$. It turns out that if we determine "the most appropriate" value in such a way that to achieve the smallest variance, this value is $p = 0$ and the corresponding "best" Generalized Hill estimator is just the Hill estimator. However if we consider the best value of $p$ as the one that leads us to the fastest rate of convergence between the distribution functions and in the sense of the Berry-Esseen theorem, the best value of $p$ is {$-1.221/\gamma$} and the estimator does not fluctuates too much if we replace this value with some value close to it.  In Section 5 we apply { the} last results on the real data example and show that in practice the difference between the last estimators is not so much important in case when the variance of the observed variable exists and it has Pareto tail with $\alpha > 2$. Therefore the extreme value theory could be applied widely in practice.

Through the paper we use the following notations: $ {\mathop{=}\limits_{}^{d}}$ is for the equality in distribution and $ {\mathop{\to}\limits_{}^{d}}$ for convergence in distribution.

\subsection{Karamata's representation and singular integrals}

In 1930 Karamata (see \cite{Karamata1930}) introduced the notion of regular
variation and proved some fundamental theorems for regularly varying (RV) functions. Here we recall his representation theorem.

\begin{thm} \textbf{Karamata's representation theorem}
 A function $U:(0,\infty)\to (0, \infty)$ is regularly varying with index $\rho$ iff
 $U$ has the representation $$U(x)=c(x)\exp(\int_1^xt^{-1}\rho(t)dt)$$
 where $\lim_{x\to \infty}c(x)=c\in (0, \infty)$ and $\lim_{t\to \infty}\rho(t)=\rho.$
\end{thm}

\begin{rmk}
One important issue which can clarify complexity of computations with concrete regularly varying tail distribution functions is the fact that albeit we have a Karamata's representation theorem (which is only an existence theorem), not always the involved  integrals are real and we can meet both complex valued functions and undefined integrals, so that not all forms of representation can be always applied. As an example may serve well Karamata's representation for $\ln (x)$ (see e.g. \cite{Resnick87}) which may need to compute integral
\begin{eqnarray}
 \int_1^t\frac{\frac{\ln x}{\ln x-1}-1}x dx=\left\{\begin{array} {ll}
undefined,&\mbox{for
$e<t$,}\\
-\pi I +\ln(\ln(t)-1), &\mbox{otherwise,}\end{array}\right.
\end{eqnarray}
where $e$ and $I$ are Euler's number and complex unit, respectively.
Karamata himself (see \cite{Karamata1963}) suggested extension of slow variation to analytic functions defined on complex plane and this was done in
\cite{Analytic}. This is important to note, since in this section we compute several limits using inverse function $b(n/k)$ for $n/k\approx \infty$ and for the sake of simplicity some complex representations of real functions will be used.
In general the  existence of such real inverses is related to  the so called asymptotic inverse functions  (see  \cite{Buldygin}).
\end{rmk}

\subsection{The second order regularly varying condition}

In this section we clarify several issues  about equivalence of asymptotic normality and  2nd order condition formulated in Theorem 4.3,
\cite{GelukdeHaan1997}. The following definition of the second order regular variation comes from \cite{deHaanStadtmueller},  \cite{deHaanFerreira}  and \cite{GelukdeHaan1997}.

\begin{df}
If the tail function of a non-negative random variable X is $\bar F := 1-F$ and
$\bar F : \mathbb{R} \to [0; 1]$  satisfies that $\bar F \in  RV_{-\alpha}$ with  $\alpha > 0.$ Then $\bar F $ is said to be of second-order regular variation with parameter $\rho \leq 0$, if there exists a function $A(t)$ that ultimately has a constant sign with $\lim_{t \to \infty} {A(t)}=0$ and a constant $c \neq 0$ such that
\begin{equation}\label{2ndregV}
\lim_{t \to \infty} \frac{\frac{\bar F(tx)}{\bar F(t)}-x^{-\alpha}}{A(t)} = H_{\alpha,\rho}(x)=cx^{-\alpha}\int_{1}^{x}u^{\rho-1}du,\ x>0
\end{equation}
Then it is written as $\bar F\in 2RV_{-\alpha, \rho}$ and $A(t)$ is  referred to as the auxiliary function of $\bar F$.
\end{df}

It is known from \cite{deHaanStadtmueller}  or a more relevant form in Geluk et
al. (1997){[14]} that if $H_{\alpha,\rho}(x)$ is not a multiple of $ x^{-\alpha}$ then $\rho < 0$ implies that there exists a $c\neq 0$  such that
$H_{\alpha,\rho}(x)=cx^{-\alpha}\frac{x^\rho-1}{\rho}$ and $|A|\in  RV_\rho$ and no other choices of $\rho$ are consistent with $ A(t)\to 0.$
There are many distributions which satisfy  the second order RV condition. These are (see e.g. \cite{Drees}): Cauchy $\gamma = 1, \rho = -2$, Fr\'echet(1) $\gamma = 1, \rho = -1$, Student t(4)       $\gamma = 1/4, \rho = -1/2$, t(10)       $\gamma = 1/10, \rho = -1/5$ or loggamma $\gamma = 1/3, \rho = 0$.

The following theorem 4.3 of \cite{GelukdeHaan1997} claims, that suppose $\bar F \in  RV_{-\alpha}$ and  Von Misses condition
(\ref{VonMisses}) holds then the asymptotic normality $N(c,\sigma^2)$, $c \not = 0$ of $\sqrt{k}(H_{X, k, n, 0} - 1/\alpha)$
 is equivalent to the second order regularly varying condition (\ref{2ndregV}).
However, it turns out that under these conditions the asymptotic normality with $c \not = 0$ is not equivalent to the 2nd order regularly varying condition.

\begin{thm} \label{Theorem4.3Geluk} (\textbf{Theorem \ref{Theorem4.3Geluk} of \cite{GelukdeHaan1997}})
Suppose $1-F\in RV_{-\alpha}$ and that the Von Misses condition holds: $F$ has density $F'$ satisfying
\begin{equation}\label{VonMisses}
\lim_{x \to \infty} \frac{x F'(x)}{\overline{F}(x)} = \alpha.
\end{equation}
Then $1-F$ is second-order regularly varying iff for some $\theta\in[0,1]$ there exists a function $U\in RV_\theta$ such that $U(t)\to \infty$
as $t\to \infty$ and there exist non-zero constants $c$ and $\sigma>0$ such that with $k=[U(n)]$ we have
$$\sqrt{k}(\hat{\gamma}_{X, k, n, 0} - \alpha^{-1})\Longrightarrow N(c,\sigma^2).$$
\end{thm}

From the proof of this theorem it is clear that they determine $c$ by the limit relation
\begin{equation}\label{c}
 c = \lim \sqrt{k}(\frac{n}{k}\int_{b(n/k)}^\infty(1-F(s))\frac{ds}{s}-\frac1{\alpha}){.}
  \end{equation}

In the following examples we show that it is possible to exist  a distribution that satisfy the conditions of this theorem and the second order regularly varying condition but not to exist subsequence of $\{n\}_{n=1}^\infty$ such that to obtain $0<c < \infty$.
We conjecture that this discrepancy  could be caused by non conformal   integral representation for
$\gamma_n$ in \cite{deHaanStadtmueller}, page 384.
Namely, between (1.8) and (1.9) in \cite{deHaanStadtmueller}, page 384
we can see \begin{equation}\label{eq}\gamma_n = \frac{n}{k}\int_{U(n/k)}^\infty ln\, s \, dF(s)\end{equation}
$$\widehat{\gamma}_n = \frac{n}{k}\int_{X_{(n-k, n)}}^\infty ln\, s \, dF_n(s)$$
$$\frac{n}{k}\int_{X_{(n-k, n)}}^\infty ln\, (s/X_{(n-k, n)}) \, dF_n(s)$$

Let us have $Pareto(1,\alpha)$ case, then $U(x)=x^{1/\alpha}$
and $1-F(x)=x^{-\alpha},x>1, \alpha>0$. Now let us denote $t:=n/k>1$
Then, according to (\ref{eq}) we have
\begin{equation}\label{eq1}
\gamma_n=t\int_{t^{1/\alpha}}^\infty \log (s) \alpha s^{-\alpha-1}ds=\alpha t \frac{1+\log (t)}{\alpha^2t}=\frac{1+\log (t)}{\alpha}\end{equation}

 Bellow (1.9) they write \begin{equation}\label{eq2} \gamma_n = \frac{n}{k}\int_{n/k}^\infty \frac{d\log U(s)}{s}=\frac{1}{\alpha}
\end{equation}

But, (\ref{eq2}) does not equal to (\ref{eq1}), and difference $\Delta_t=\frac{\log (t)}{\alpha}$ between both integral representations of $\gamma_n$  converge to $\infty$ for $t=n/k\to \infty$.

We guess, that non-conformal integral  representation could be  caused by mismatch  of Theorem 4.1 from \cite{DavisResnick}, which is cited just before introducing representation (\ref{eq}) in \cite{deHaanStadtmueller}, since if we calculate the expression in \cite{DavisResnick} in Pareto case it gives a correct value, namely
$$a^*(t) = t \int_{ln\, U(t)}^\infty \overline{F}(e^s) ds = t \int_{\frac{1}{\alpha}ln\, t}^\infty e^{-\alpha s} ds = -\frac{t}{\alpha} e^{-\alpha y}|_{\frac{1}{\alpha}ln\, t}^\infty = \frac{1}{\alpha}.$$

Thus in the following  Examples 1-3 we will also show that $c=\infty$ is accompanied by infinite difference between above representations (\ref{eq}) and (\ref{eq1}) for $\gamma_n$  when $n/k\to \infty.$ We will denote this difference $\Delta_n:=\gamma_n(\ref{eq})-\gamma_n(\ref{eq1}),$ where
$\gamma_n(\ref{eq}),\gamma_n(\ref{eq1})$ denote the $\gamma_n$ from above representations (\ref{eq}) and (\ref{eq1}), respectively.

\begin{ex}  (\textbf{Hall/Weiss class})
The most common example of $\bar F \in  RV_{-\alpha}$, $\alpha > 0$, $\rho < 0$, is the so called Hall/Weiss class of distributions
\begin{equation}\label{HallWeiss}F(x) = 1 - \frac{1 + x^\rho}{2x^\alpha}, \quad x > 1.
\end{equation}
Briefly we will denote this by $F \in HW(\alpha, \rho)$.

$\bar F \in  2RV_{-\alpha, \rho}$ because

$$\frac{\overline{F}(tx)}{\overline{F}(t)} - x^\alpha = x^{-\alpha} \frac{t^\rho}{1 + t^\rho} (x^\rho - 1),$$
 we can chose $A(t) = \frac{\rho t^\rho}{1 + t^\rho}$ and this function is always negative and regularly varying with parameter $\rho$.
The distributions from this type also satisfy the Von Misses condition (\ref{VonMisses}).
$$\lim_{x \to \infty} \frac{x F'(x)}{\overline{F}(x)} = \lim_{x \to \infty} \frac{\alpha + (\alpha - \rho)x^\rho}{1 + x^\rho} = \alpha.$$

Now let us try to find subsequence of $n \in \mathbb{N}$ and constant $c \not = 0$ and $c < \infty$, described in (\ref{c}) such that to apply Theorem \ref{Theorem4.3Geluk}.

\textbf{Case 1: Hall/Weiss distribution  $HW(1, -1)$}

 Let us consider $F \in HW(1, -1)$, i.e.
$$F(x) = 1 - \frac{1 + x^{-1}}{2x}, \quad x > 1.$$
We already mentioned that the tail of this distribution function belongs to $RV_{-1} \cup  2RV_{1, -1}$ and it satisfies the Von Misses condition (\ref{VonMisses}). It is not difficult to calculate that
$$\frac{1}{\overline{F}}(x) = \frac{2x^2}{x + 1}, \quad x > 1$$
Therefore
$$b(p) = \left(\frac{1}{\overline{F}}\right)^\leftarrow(p) = \frac{p + \sqrt{p^2 + 8p}}{4}, \quad p  > 1.$$
Let us now find $c$ in (\ref{c}).
$$c = \lim_{n \to \infty} \sqrt{o(n)}(\frac{n}{o(n)}\int_{b(n/o(n))}^\infty(1-F(s))\frac{ds}{s}-\frac {1}{\alpha}) = \lim_{n \to \infty} \sqrt{o(n)}\left(\frac{n}{o(n)}\int_{b(n/o(n))}^\infty \frac{1 + s^{-1}}{2s^2}ds-1\right) =$$
$$= \lim_{n \to \infty} \sqrt{o(n)}\left[\frac{n}{2o(n)}\left(\frac{1}{b(n/o(n))} + \frac{1}{2b^2(n/o(n))}\right)-1\right] =$$
$$= \lim_{n \to \infty} \sqrt{o(n)}\left[\frac{2}{1 + \sqrt{1 + \frac{8}{n/o(n)}}} + \frac{4}{n/o(n)[1 + \sqrt{1 + \frac{8}{n/o(n)}}]^2}-1\right] = \lim_{n \to \infty} \sqrt{o(n)} = \infty.$$
Therefore such a subsequence, $k(n)$ mentioned in Theorem \ref{Theorem4.3Geluk} does not exist.

Now, let us compute the difference $\Delta_n:=\gamma_n(\ref{eq})-\gamma_n(\ref{eq1}), t:=n/k>0$
We have $$\gamma_t(\ref{eq})= -t\frac{2\ln b(t) +1+2b(t)+2b(t)\ln b(t)}{4b(t)^2}$$
$$\gamma_t(\ref{eq1})= \frac{\sqrt{t(8+t)}+4-t}{8}$$
and finally
$\lim_{n/k\to \infty}\Delta_n=\lim_{t\to \infty}\left(\gamma_t(\ref{eq})-\gamma_t(\ref{eq1})\right)=\infty.$

\textbf{Case 2: Hall/Weiss distribution  $HW(2, -1)$}

 For $F \in HW(2, -1)$ we have
$$F(x) = 1 - \frac{1 + x^{-1}}{2x^2}, \quad x > 1.$$
The tail of this c.d.f. belongs to $RV_{-1} \cup  2RV_{2, -1}$ and it satisfies the Von Misses condition (\ref{VonMisses}).
$$\frac{1}{\overline{F}}(x) = \frac{2x^3}{x + 1}, \quad x > 1$$
{Therefore for $p > 0$
\begin{equation}\label{HallWeissBP}
b(p) =  \left(\frac{1}{\overline{F}}\right)^\leftarrow(p) = \left(\frac{2x^3}{x + 1}\right)^\leftarrow(p)=\frac{\sqrt[3]{54p+6\sqrt{-6p^3+81p^2}}}6+\frac{p}{\sqrt[3]{54p+6\sqrt{-6p^3+81p^2}}},
\end{equation}}

We have to compute $$c = \lim_{n \to \infty} \sqrt{o(n)}(\frac{n}{o(n)}\int_{b(n/o(n))}^\infty(1-F(s))\frac{ds}{s}-\frac {1}{\alpha}) = \lim_{n \to \infty} \sqrt{o(n)}\left(\frac{n}{o(n)}\int_{b(n/o(n))}^\infty \frac{1/s + 1}{2s^3}ds-\frac {1}{2}\right).$$

For further purpose, let us denote $n/o(n):=t,$ we know $n/o(n)\to \infty$ for all $o(n),$ thus we study $t\to \infty$.

We have for all $$b(n/o(n))>0:\ t\int_{b(t)}^\infty \frac{1/s + 1}{2s^3}ds=t\frac{2+3b(t)}{12b(t)^3}=$$

\begin{equation}\label{HallWeissBPexpr}
=\frac{t(2+\frac{\sqrt[3]{54t+6\sqrt{-6t^3+81t^2}}}{2}+\frac{3t}{\sqrt[3]{54t+6\sqrt{-6t^3+81t^2}}})}{\frac{\sqrt[3]{54t+6\sqrt{-6t^3+81t^2}}}{6}
+\frac{t}{\sqrt[3]{54t+6\sqrt{-6t^3+81t^2}}}}
\end{equation}

By using of the last expression (\ref{HallWeissBPexpr}) we  have
$$\lim_{n/o(n)\to \infty} t\int_{b(t)}^\infty \frac{1/s + 1}{2s^3}ds=6$$

Thus,{ for any choice of $k(n) = o(n)$ we obtain that}  $$c = \lim_{k=o(n), k\to \infty, n\to \infty} \sqrt{o(n)}(\frac{n}{o(n)}\int_{b(n/o(n))}^\infty(1-F(s))\frac{ds}{s}-\frac {1}{\alpha})=+\infty$$

Now, let us compute the difference $\Delta_n:=\gamma_n(\ref{eq})-\gamma_n(\ref{eq1}), t:=n/k>0$
We have $$\gamma_t(\ref{eq})= t\frac{6\ln b(t) +2+3b(t)+6b(t)\ln b(t)}{12b(t)^3}$$
For $\gamma_t(\ref{eq1}),$ the integral was not able to write in a form of elementary function,
so we shall compute only the limit, necessary for limiting difference,
i.e. $$\lim_{t\to \infty}\gamma_t(\ref{eq1})=\lim_{t\to \infty}t\frac{d\ln b(t)}{dt}=\frac12,$$
and finally
$\lim_{n/k\to \infty}\Delta_n=\lim_{t\to \infty}\left(\gamma_t(\ref{eq})-\gamma_t(\ref{eq1})\right)=\infty.$

\textbf{Case 3: Hall/Weiss distribution  $HW(1, -2)$}

 For $F \in HW(1, -2)$ we have
$$F(x) = 1 - \frac{1 + x^{-2}}{2x}, \quad x > 1.$$
As a particular case of Hall/Weiss distribution  the tail of this c.d.f. belongs to $RV_{-1} \cup  2RV_{1, -2}$ and it satisfies the Von Misses condition (\ref{VonMisses}). Let us now try to calculate $c$.
$$\frac{1}{\overline{F}}(x) = \frac{2x^3}{x^2 + 1}, \quad x > 1$$
Therefore for $p > 1$
$$b(p) = \left(\frac{1}{\overline{F}}\right)^\leftarrow(p) = \frac{\sqrt[3]{54p+p^3+6\sqrt{81p^2+3p^4}}}{6}+\frac{p^2}{6\sqrt[3]{54p+p^3+6\sqrt{81p^2+3p^4}}}+\frac{y}{6}. $$

We have to compute (\ref{c})
$$c = \lim_{n \to \infty} \sqrt{o(n)}(\frac{n}{o(n)}\int_{b(n/o(n))}^\infty(1-F(s))\frac{ds}{s}-\frac {1}{\alpha}) = \lim_{n \to \infty} \sqrt{o(n)}\left(\frac{n}{o(n)}\int_{b(n/o(n))}^\infty \frac{s^2 + 1}{2s^4}ds-1\right).$$

We have for all $$b(n/o(n))>0:\ t\int_{b(t)}^\infty \frac{s^2 + 1}{2s^4}ds=t\frac{1+3b(t)^2}{6b(t)^3}=$$

\begin{equation} \label{HallWeissBPexpr2}
=\frac{t(1+3(\frac{\sqrt[3]{54t+t^3+6\sqrt{81t^2+3t^4}}}{6}+\frac{t^2}{6\sqrt[3]{54t+t^3+6\sqrt{81t^2+3t^4}}}+\frac{t}6)^2)}{(\frac{\sqrt[3]{54t+t^3+6\sqrt{81t^2+3t^4}}}{6}
+\frac{t^2}{6\sqrt[3]{54t+t^3+6\sqrt{-81t^2+3t^4}}}+\frac{t}6)^3}
\end{equation}

By using of the last expression (\ref{HallWeissBPexpr2}) we  have
$$\lim_{n/o(n)\to \infty} t\int_{b(t)}^\infty \frac{s^2 + 1}{2s^4}ds=6$$

Thus, { for any choice of $k(n) = o(n)$ we obtain that} $$c = \lim_{k=o(n), k\to \infty, n\to \infty} \sqrt{o(n)}(\frac{n}{o(n)}\int_{b(n/o(n))}^\infty(1-F(s))\frac{ds}{s}-\frac {1}{\alpha})=+\infty$$

Now, let us compute the difference $\Delta_n:=\gamma_n(\ref{eq})-\gamma_n(\ref{eq1}), t:=n/k>0$
We have $$\gamma_t(\ref{eq})= t\frac{1+3b(t)^2\ln b(t) +3\ln b(t)+3b(t)^2}{6b(t)^3}$$
For $\gamma_t(\ref{eq1}),$ the integral was not able to be written in a form of elementary function,
so we shall compute only limit, necessary for limiting difference,
i.e. $$\lim_{t\to \infty}\gamma_t(\ref{eq1})=\lim_{t\to \infty}t\frac{d\ln b(t)}{dt}=1,$$
and finally
$\lim_{n/k\to \infty}\Delta_n=\lim_{t\to \infty}\left(\gamma_t(\ref{eq})-\gamma_t(\ref{eq1})\right)=\infty.$

\end{ex}

\begin{ex} \textbf{Log Erlang(2,1) }

We have $\rho = 0$, $\alpha = 1$ (see \cite{GelukdeHaan1997}).
$$F(x) = 1 - \frac{1 + \ln\, x}{x}, \quad x > 1$$
$$\frac{1}{\overline{F}}(x) = \frac{x}{1+\ln(x)}$$
$$b(p) = \left(\frac{1}{\overline{F}}\right)^\leftarrow(p) = \exp(-LW(-\frac{1}{ep})-1),$$
where $LW$ is the principal  real valued branch of Lambert W function, see \cite{Stehlik03}.
The slowly varying function in this case is $L(x) = 1 + \ln\, x.$
We have
$$\lim_{t\to\infty}t\int_{b(t)}^\infty (1-F(s))\frac{ds}{s}=\lim_{t\to\infty}t\int_{b(t)}^\infty \frac{1+\ln(s)}{s^2}ds=$$
$$=e\lim_{t\to\infty}\frac{-1+LW(-\frac{1}{et})}{LW(\frac{-1}{et})}=\infty$$.

Let us choose $k=o(n)$, then  we have
$$c = \lim_{n\to\infty} \sqrt{o(n)}(\frac{n}{o(n)}\int_{b(\frac{n}{o(n)})}^\infty(1-F(s))\frac{ds}{s}-\frac {1}{\alpha}) = \infty. $$

Now, let us compute the difference $\Delta_n:=\gamma_n(\ref{eq})-\gamma_n(\ref{eq1}), t:=n/k>0$
We have $$\gamma_t(\ref{eq})= -\frac{LW(-\frac{1}{et})^2+1}{LW(-\frac{1}{et})}$$
$$\gamma_t(\ref{eq1})= -\frac{1-LW(-\frac{1}{et})+etLW(-\frac{1}{et})}{LW(-\frac{1}{et})}$$
and finally
$\lim_{n/k\to \infty}\Delta_n=\lim_{t\to \infty}\left(\gamma_t(\ref{eq})-\gamma_t(\ref{eq1})\right)=\infty.$

\end{ex}

\begin{ex} \textbf{Slowly varying function satisfying  2nd order RV condition with $\rho=0$}
$$F(x) =  1 - ex^{-{1}}\ln(x), \quad x > e$$
It is easy to show that this function satisfies 2nd order RV condition with $\rho=0$ with
$A(t)=(\ln (t))^{-1}$.
We have
$$\frac{1}{\overline{F}}(x) = \frac{x}{e\ln(x)}$$
$$b(p) = \left(\frac{1}{\overline{F}}\right)^\leftarrow(p) = -epLW(-\frac{1}{ep}),$$
here $LW$ is  the principal  real valued branch of Lambert W function, see \cite{Stehlik03}.
We have
$$\lim_{t\to\infty}t\int_{b(t)}^\infty (1-F(s))\frac{ds}{s}=\lim_{t\to\infty}t\int_{b(t)}^\infty \frac{e\ln(s)}{s^2}ds=$$
$$=e\lim_{t\to\infty}t\frac{1+\ln(-etLW(-\frac{1}{et}))}{-etLW(-\frac{1}{et})}=\infty$$.
Let us choose $k=o(n)$, then  we have
$$c = \lim_{n\to\infty} \sqrt{o(n)}(\frac{n}{o(n)}\int_{b(\frac{n}{o(n)})}^\infty(1-F(s))\frac{ds}{s}-\frac {1}{\alpha}) = \infty. $$

Now, let us compute the difference $\Delta_n:=\gamma_n(\ref{eq})-\gamma_n(\ref{eq1}), t:=n/k>0$
We have $$\gamma_t(\ref{eq})= -\frac{-LW(-\frac{1}{et})+1+LW(-\frac{1}{et})^2}{LW(-\frac{1}{et})}$$
$$\gamma_t(\ref{eq1})= -\frac{1-LW(-\frac{1}{et})+etLW(-\frac{1}{et})}{LW(-\frac{1}{et})}$$
and finally
$\lim_{n/k\to \infty}\Delta_n=\lim_{t\to \infty}\left(\gamma_t(\ref{eq})-\gamma_t(\ref{eq1})\right)=\infty.$

\end{ex}

Moreover,  $F(x) = 1 - (x/\delta)^{-\alpha}$, $x > \delta$ that is $Pareto(\alpha, \delta)$ distribution{,} belongs to $RV_{-\alpha}${. I}t does not satisfy the second order regularly varying condition and we have $c = 0$.

{ Further on we  clarify the conditions that we need to impose in order to obtain asymptotic normality of the Generalized Hill and in particular of the Hill estimator. Some alternative approaches, albeit preliminary, could be find in \cite{HeauslerTeugels} and \cite{Pancheva}. Therefore we  show that asymptotic normality is possible to be achieved without the second order regularly varying condition. }

\section{The limiting distribution of the Generalized Hill estimator for fixed number of order statistics}

\subsection{Pareto case}

We start our investigations with the case when the observed random variable is Pareto distributed and find the exact distribution of the Generalized Hill estimators for fixed number of order statistics $k$ which is less than  the sample size $n$.

{\begin{prop} If the c.d.f. $F$ is Pareto($\alpha, \delta$), i.e. if
\begin{equation}\label{Pareto}
 F(x) = \left\{
                  \begin{array}{ccc}
                    0 & , & x < \delta \\
                    1 - \left(\frac{\delta}{x}\right)^{\alpha} & , & x \geq \delta
                  \end{array}
                \right .,
\end{equation}
then
\begin{equation}\label{Huniform_def}
H_{X, k, n, p} {\mathop{=}\limits_{}^{d}} \frac{1}{k}\sum_{i=1}^{k}U_i^{-p/\alpha}, \quad p \in \mathbb{R}, \quad p \not = 0,
\end{equation}
{ $$\widehat{\gamma}_{X, k, n, p}  \,{\mathop{=}\limits_{}^{d}}\, \frac{1}{p}\left(1- \left[\frac{1}{k}\sum_{i=1}^{k}\,U_i^{-\frac{p}{\alpha }} \right]^{-1} \right)   \quad p \in \mathbb{R}, \quad p \not= 0.$$}

$$\hat{\gamma}_{X, k, n, 0} \,{\mathop{=}\limits_{}^{d}}\, \frac{1}{k}\sum_{i=1}^{k}\ln \,U_i^{-\frac{1}{\alpha }}, $$
where $U_1, U_2, ..., U_n$ are i.i.d.
uniformly distributed r.v's on $(0, 1)$.
\end{prop}}

{\bf Proof:} Let
$\mathbf{U}_{(1,n)} \leq \mathbf{U}_{(2 ,n)} \leq ... \leq
\mathbf{U}_{(n ,n)}$ be the upper order statistics of $U_1, U_2, ..., U_n$.

It is not difficult to check that
\begin{equation}\label{Prop3_1}
\left\{ 1 - U_{(n-i+1,n)}, i = 1, 2, ..., n\right\} {\mathop{=}\limits_{}^{d}} \left\{U_{(i, n)}, i = 1, 2, ..., n\right\}
\end{equation}
and
\begin{equation}\label{Prop3_2}
    \left\{ \frac{U_{(i,n)}}{U_{(k+1,n)}}, i = 1, 2, ..., k\right\} {\mathop{=}\limits_{}^{d}} \left\{U_{(i,k)}, i = 1, 2, ..., k\right\}.
\end{equation}

Recall, the probability quantile transform states that
\begin{equation}\label{PropIntTr}
\left\{ X_{(i,n)},  i = 1, 2, ... , n \right\}
{\mathop{=}\limits_{}^{d}}\,
\left\{ F^\leftarrow (U_{(i,n)}) ,  i = 1, 2, ... , n \right\},  i =
1, 2, ... , n.
\end{equation}

Therefore for $F$ - Pareto
\begin{equation}\label{PropIntTr}
\left\{ X_{(i,n)},  i = 1, 2, ... , n \right\}
{\mathop{=}\limits_{}^{d}}\,
\left\{\delta  (1 - \mathbf{U}_{(i,n)})^{-1/\alpha},  i =
1, 2, ... , n \right\}.
\end{equation}

For $p \not= 0$, the definition (\ref{HXknp}) and equalities (\ref{Prop3_1}) and
(\ref{Prop3_2}) give
$$H_{X, k, n, p} = \frac{1}{k}\sum_{i=1}^{k}\left(\frac{{X}_{(n-i+1, n)}}{X_{(n - k,n)}}\right)^{p} \,{\mathop{=}\limits_{}^{d}} \frac{1}{k}\sum_{i=1}^{k}\left(\frac{1 - U_{(n-i+1, n)}}{1 - U_{(n - k, n)}}\right)^{-\frac{p}{\alpha}}  \,{\mathop{=}\limits_{}^{d}} \frac{1}{k}\sum_{i=1}^{k}\left(\frac{\mathbf{U}_{(i, n)}}{\mathbf{U}_{(k + 1, n)}}\right)^{-\frac{p}{\alpha}}$$
$${\mathop{=}\limits_{}^{d}}  \frac{1}{k}\sum_{i=1}^{k}\mathbf{U}_{(i, k)}^{-\frac{p}{\alpha}}\,{\mathop{=}\limits_{}^{d}} \, \frac{1}{k}\sum_{i=1}^{k}\mathbf{U}_{ i}^{-\frac{p}{\alpha}}.$$
Analogously
$$\hat{\gamma}_{X, k, n, 0} = \frac{1}{k}\sum_{i=1}^{k}\ln\, \left(\frac{{X}_{(n-i+1, n)}}{X_{(n - k,n)}}\right) \,{\mathop{=}\limits_{}^{d}} \frac{1}{k}\sum_{i=1}^{k}\ln\,\left(\frac{1 - U_{(n-i+1, n)}}{1 - U_{(n - k, n)}}\right)^{-\frac{1}{\alpha }}  \,{\mathop{=}\limits_{}^{d}} \frac{1}{k}\sum_{i=1}^{k}\ln\,\left(\frac{\mathbf{U}_{(i, n)}}{\mathbf{U}_{(k + 1, n)}}\right)^{-\frac{1}{\alpha }}$$
$${\mathop{=}\limits_{}^{d}}  \frac{1}{k}\sum_{i=1}^{k}\ln\, \mathbf{U}_{(i, k)}^{-\frac{1}{\alpha }}\,{\mathop{=}\limits_{}^{d}} \, \frac{1}{k}\sum_{i=1}^{k} \ln\,\mathbf{U}_{ i}^{-\frac{1}{\alpha }}.$$
\hfill $\Box$

{\it Note:}  In this case:

1.  the distribution of $\hat{\gamma}_{X, k, n, 0}$ is $Gamma(k, k\alpha )$. It is well known that it has mean $1/\alpha$ and variance $1/(k\alpha^2)$. Therefore the larger the number of order statistics, the smaller the variance of the limiting distribution.

2.  $H_{X, k, n, -\alpha}$ is Irwin - Hall distributed.

\subsection{Regularly varying case}

In the next statement we suppose that the sample size increases and do not suppose the exact Pareto distribution
of the observed random variable, but only regularly varying tail of its distribution function and obtain the same limit distribution of the Generalized Hill estimator for fixed number of order statistics, $k$.

{\begin{prop} Let $\mathbf{X}_1, \mathbf{X}_2, ..., \mathbf{X}_n$
be independent copies of $\mathbf{X}$ with d.f. $F$,
\begin{equation}\label{RV_cond}
\bar{F} \in
RV_{-\alpha}.
\end{equation}
For fixed $k, \alpha$ and $p < 0$ and $n \to \infty$
\begin{equation}\label{GenHillT}
 H_{X, k, n, p}\, {\mathop{\to}\limits_{}^{d}}\, \frac{1}{k}\sum_{i=1}^{k} \mathbf{U}_i ^{-\frac{p}{\alpha}} ,
 \end{equation}
 { $$\widehat{\gamma}_{X, k, n, p}  \, {\mathop{\to}\limits_{}^{d}}\, \frac{1}{p}\left(1- \left[\frac{1}{k}\sum_{i=1}^{k}\,U_i^{-\frac{p}{\alpha}} \right]^{-1} \right)   \quad p \in \mathbb{R}, \quad p \not= 0.$$}
\begin{equation}\label{HillT}
\hat{\gamma}_{X, k, n, 0}\, {\mathop{\to}\limits_{}^{d}}\, \frac{1}{k}\sum_{i=1}^{k}\ln \,U_i^{-\frac{1}{\alpha }},
\end{equation}
where $\mathbf{U}_1, \mathbf{U}_2, ..., \mathbf{U}_k$ are i.i.d.
uniformly distributed r.v's on $(0, 1)$.
\end{prop}

{\bf Proof:} The prove of (\ref{HillT}) follow immediately from Theorem 1 in \cite{HeauslerTeugels}, the quantile transformation and the presentation of the Erlang distributed random variable as sum of i.i.d. exponentially distributed random variables.

Let us now prove (\ref{GenHillT}).

Denote the distribution of $\mathbf{X}_1^p, \mathbf{X}_2^p, ..., \mathbf{X}_n^p$ by $F_p$, then
$$\overline{F}_p(x) = 1 - F_p(x) = P(\mathbf{X}_1^p > x) = P(\mathbf{X}_1 > x^\frac{1}{p}) = \overline{F}(x^\frac{1}{p}).$$

(\ref{RV_cond}) imply that $\overline{F}_p(x) \in RV_{-\frac{\alpha}{p}}.$

By the probability quantile transformation, (\ref{Prop3_1})  and (\ref{RV_cond}) we have that there exists a slowly varying function $L$ such that
$$\left\{ \mathbf{X}_{(i, n)}^p,  i = 1, 2, ... , n \right\} {\mathop{=}\limits_{}^{d}}\, \left\{ F_p^\leftarrow (\mathbf{U}_{(i, n)}),  i = 1, 2, ... , n \right\} $$
$${\mathop{=}\limits_{}^{d}}\,\, \left\{ F_p^\leftarrow (1 - \mathbf{U}_{(n - i + 1, n)}),  i = 1, 2, ... , n \right\} $$
$${\mathop{=}\limits_{}^{d}}\,\, \left \{\left (\frac{1}{\overline{F_p}}\right)^\leftarrow \left ( \frac{1}{\mathbf{U}_{(n - i + 1, n)}}\right),  i = 1, 2, ... , n  \right \} $$
$$= \left \{ \left( \frac{1}{\mathbf{U}_{(n - i + 1, n)}}\right )^{\frac{p}{\alpha}}L\left ( \frac{1}{\mathbf{U}_{(n - i + 1, n)}}\right ), \quad i = 1, 2, ... , n \right \}.$$

The reciprocal of a
uniformly distributed r.v. is a.s. greater than one. The Karamata-representation theorem for regularly varying functions entails that there exist measurable and bounded functions $c(x)$, converging to a constant and  $\varepsilon(x)$, converging to $0$, when the argument is close to infinity, and $B > 0$, such that
 $$\left\{ \mathbf{X}_{(i, n)}^p,  i = 1, 2, ... , n \right\} {\mathop{=}\limits_{}^{d}}\,
\left\{\mathbf{U}_{(n - i + 1,n)}^{-\frac{p}{\alpha}} c \left(\frac{1}{\mathbf{U}_{(n - i + 1,n)}} \right)exp\left(\int_{B}^{\mathbf{U}_{(n - i + 1,n)}^{-1}} \frac{\varepsilon(x)}{x}\,dx\right) ,  i = 1, 2, ... , n \right\} $$

Consider $H_{X, k, n, p}$ defined in (\ref{HXknp}).
$$H_{X, k, n, p} = \frac{1}{k}\sum_{i=1}^{k} \left ( \frac{\mathbf{U}_{(i,n)}}{\mathbf{U}_{(k + 1,n)}} \right)^{-\frac{p}{\alpha}}\frac{c \left(\frac{1}{\mathbf{U}_{(i,n)}} \right)}{c \left(\frac{1}{\mathbf{U}_{(k + 1,n)}}\right)}exp\left(\int_{\mathbf{U}_{(k + 1,n)}^{-1}}^{\mathbf{U}_{(i,n)}^{-1}} \frac{\varepsilon(x)}{x}\,dx\right),$$
where $c(x) \to c_0 \in (0, \infty)$ as $x \to \infty$, $\varepsilon:
R^+ \to R^+$ and $\varepsilon(t) \to 0$ as $t \to \infty$.

By (\ref{Prop3_2}), for $n \in \mathbb{N}$ and $k = 1, 2, ..., n$,
$$ \frac{1}{k}\sum_{i=1}^{k}\left(\frac{\mathbf{U}_{(i ,n)}}{\mathbf{U}_{(k + 1,n)}}\right)^{-\frac{p}{\alpha}} \, {\mathop{=}\limits_{}^{d}} \, \frac{1}{k}\sum_{i=1}^{k}\mathbf{U}_{(i,k)}^{-\frac{p}{\alpha}} =  \frac{1}{k}\sum_{i=1}^{k}\mathbf{U}_{ i}^{-\frac{p}{\alpha}}.$$

If
\begin{equation}\label{Bil_prob}
\Delta_n : = \left | H_{X, k, n, p} -
\frac{1}{k}\sum_{i=1}^{k}\left(\frac{\mathbf{U}_{(i
,n)}}{\mathbf{U}_{(k + 1,n)}}\right)^{- \frac{p}{\alpha}} \right |
{\mathop{\rightarrow}\limits_{}^{P}}\, 0,
\end{equation}
then in distribution $\lim_{n \to \infty} H_{X, k, n, p}  = \lim_{n \to
\infty} \frac{1}{k}\sum_{i=1}^{k}\left(\frac{\mathbf{U}_{( i
,n)}}{\mathbf{U}_{(k + 1,n)}}\right)^{-\frac{p}{\alpha}}$ (cf.
Theorem 4.1. of \cite{Billingsley77}) and the proof would be completed.

We check (\ref{Bil_prob}). By the triangle inequality
$$ \Delta_n \leq  \frac{1}{k}\sum_{i=1}^{k} \left ( \frac{\mathbf{U}_{(i ,n)}}{\mathbf{U}_{(k + 1,n)}} \right)^{-\frac{p}{\alpha}}  \left | \frac{c \left(\frac{1}{\mathbf{U}_{(i,n)}} \right)}{c \left(\frac{1}{\mathbf{U}_{(k + 1,n)}}\right)}exp\left(\int_{\mathbf{U}_{(k + 1,n)}^{-1}}^{\mathbf{U}_{(i,n)}^{-1}} \frac{\varepsilon(x)}{x}\,dx\right) - 1 \right | .$$

$p \leq 0$ imply that for $n \in \mathbb{N}$, $ k = 1, 2, ..., n$ and $i = 1, 2, ..., k$,
$0 \leq \left ( \frac{\mathbf{U}_{(i,n)}}{\mathbf{U}_{(
k+1,n)}} \right)^{-\frac{p}{\alpha}} \leq 1.$ Thus
$$0 \leq \Delta_n \leq \frac{1}{k}\sum_{i=1}^{k}\left | \frac{c \left(\frac{1}{\mathbf{U}_{(i,n)}} \right)}{c \left(\frac{1}{\mathbf{U}_{(k + 1 ,n)}}\right)}exp\left(\int_{\mathbf{U}_{(k+1,n)}^{-1}}^{\mathbf{U}_{(i,n)}^{-1}} \frac{\varepsilon(x)}{x}\,dx\right) - 1 \right |$$
Now we have to show that  the summands in the
above expression converge in probability to zero for $n \to \infty$. The function $T(x,
y) = x.y$ is continuous in the point $(x, y) = (1, 1)$. In order to
use the continuity of composition we have to check the following two
convergences
\begin{equation}\label{c_over_c}
    \frac{c \left \{\frac{1}{\mathbf{U}_{(i,n)}} \right\}}{c \left \{\frac{1}{\mathbf{U}_{(k + 1,n)}}\right\}} {\mathop{\rightarrow}\limits_{}^{\mathbb{P}}} 1
\end{equation}
and
\begin{equation}\label{e_na}
exp\left \{\int_{\mathbf{U}_{(k + 1, n)}^{-1}}^{\mathbf{U}_{(i, n)}^{-1}} \frac{\varepsilon(x)}{x}\,dx\right\} \, {\mathop{\rightarrow}\limits_{}^{\mathbb{P}}}\, 1, \quad n \to \infty.
\end{equation}

The function $c: R^+ \to R^+$ is such that $c(x) \to c_0 \in (0,
\infty)$ as $x \to \infty$. Recall $\frac{\mathbf{U}_{(i ,n)}}{\frac{i}{n}} {\mathop{\rightarrow}\limits_{}^{a.s.}}\,1$  $n \to \infty$. Then
$\mathbf{U}_{(i ,n)}
{\mathop{\rightarrow}\limits_{}^{a.s.}}\, 0$ and $\mathbf{U}_{(
k + 1,n)} {\mathop{\rightarrow}\limits_{}^{a.s.}}\, 0$ as $n \to
\infty$. Thus, (\ref{c_over_c}) follows by continuity of $g(x, y) =
\frac{x}{y}$ in $(x, y) = (c_0, c_0)$ and the Slutsky theorem (about the continuity in probability of the composition).

Consider (\ref{e_na}). It
is enough to prove that
$$\int_{\mathbf{U}_{(k + 1,n)}^{-1}}^{\mathbf{U}_{(i ,n)}^{-1}} \frac{\varepsilon(x)}{x}\,dx \, {\mathop{\rightarrow}\limits_{}^{\mathbb{P}}}\,\,  0.$$

By $\frac{\mathbf{U}_{(i ,n)}}{\frac{i}{n}} {\mathop{\rightarrow}\limits_{}^{a.s.}}\,1$,  $n \to \infty$ we have
$\mathbf{U}_{(i ,n)}
{\mathop{\rightarrow}\limits_{}^{a.s.}}\, 0$ and $\mathbf{U}_{(
k + 1,n)} {\mathop{\rightarrow}\limits_{}^{a.s.}}\, 0$ as $n \to
\infty$.  In view of Karamata-representation for regularly varying functions
$\varepsilon(t) \to 0$ as $t \to \infty$.  Consequently a.s. for
$\varepsilon_0 > 0$ there exists $n_{\varepsilon_0} \in \mathbb{N}$,
such that for $n > n_{\varepsilon_0}$,
$$|\int_{\mathbf{U}_{(k + 1,n)}^{-1}}^{\mathbf{U}_{(i ,n)}^{-1}} \frac{\varepsilon(x)}{x}\,dx| \leq \varepsilon_0 |\int_{\mathbf{U}_{(k + 1,n)}^{-1}}^{\mathbf{U}_{(i ,n)}^{-1}} \frac{1}{x}\,dx | = \varepsilon_0\,|\,\ln \, \frac{\mathbf{U}_{(k + 1,n)}}{\mathbf{U}_{(i ,n)}} \,|.$$

Let $\epsilon > 0$. By (\ref{Prop3_2})
$$0 \leq P( |\int_{\mathbf{U}_{(k + 1,n)}^{-1}}^{\mathbf{U}_{(i ,n)}^{-1}} \frac{\varepsilon(x)}{x}\,dx| \geq \epsilon) $$
$$\leq P( \varepsilon_0\,|\,\ln \, \frac{\mathbf{U}_{(k + 1,n)}}{\mathbf{U}_{(i ,n)}} \,| \geq \epsilon) \leq P( \varepsilon_0\, | \,\ln \, \mathbf{U}_{(i ,k)}| \geq \epsilon).$$

The random variable $\ln \, \mathbf{U}_{(i, k)}$ does not
depend on $n$ and it is a.s. finite, hence  for $\varepsilon_0 \to
0$ we obtain that (\ref{e_na}) is satisfied and we complete the proof.

 \hfill $\Box$

\subsection{Diagnostic plots for fixed $k$}

 For applications it is very important to  consider the behavior of these estimators for fixed sample size $n$ and fixed number of order statistics $k$.}
A very frequently used example, that shows the disadvantages and mainly the slow rate of convergence of the Hill estimator is the distribution with the following quantile function
\begin{equation}\label{HillHorror}
F^{\leftarrow}(p)=-(1-p)^{-1/\alpha}\ln (1-p), \,\,p\in (0,1).
\end{equation}
See \cite{EMK}. We call this distribution Hill-horror distribution with parameter $\alpha > 0$. Briefly {$F \in HillH(\alpha)$.
 We can determine easily
$U(n)=F^{\leftarrow}(1-1/n)=n^{1/\alpha}\ln n$
and this distribution satisfies the 2nd order regular variation condition with $A(t) = 1/\ln\, t$ and $\rho = 0$.}

\cite{EMK} show that in that case it is almost impossible to determine $\alpha$ using Hill plot. Therefore they call this plot "Hill horror plot".

In the next figures we show that when fix $k$ "sufficiently big" and let $n > k$ to infinity, using Proposition 2 we could determine $\alpha$ with reasonably small error also for this Hill horror distribution. We suppose that we do not know the exact distribution therefore we do not use the whole sample, but only follow the algorithm that allows us to use the above theorems.

\begin{figure}
\begin{center}
\begin{minipage}[t]{0.45\linewidth}
    \includegraphics[scale=.3]{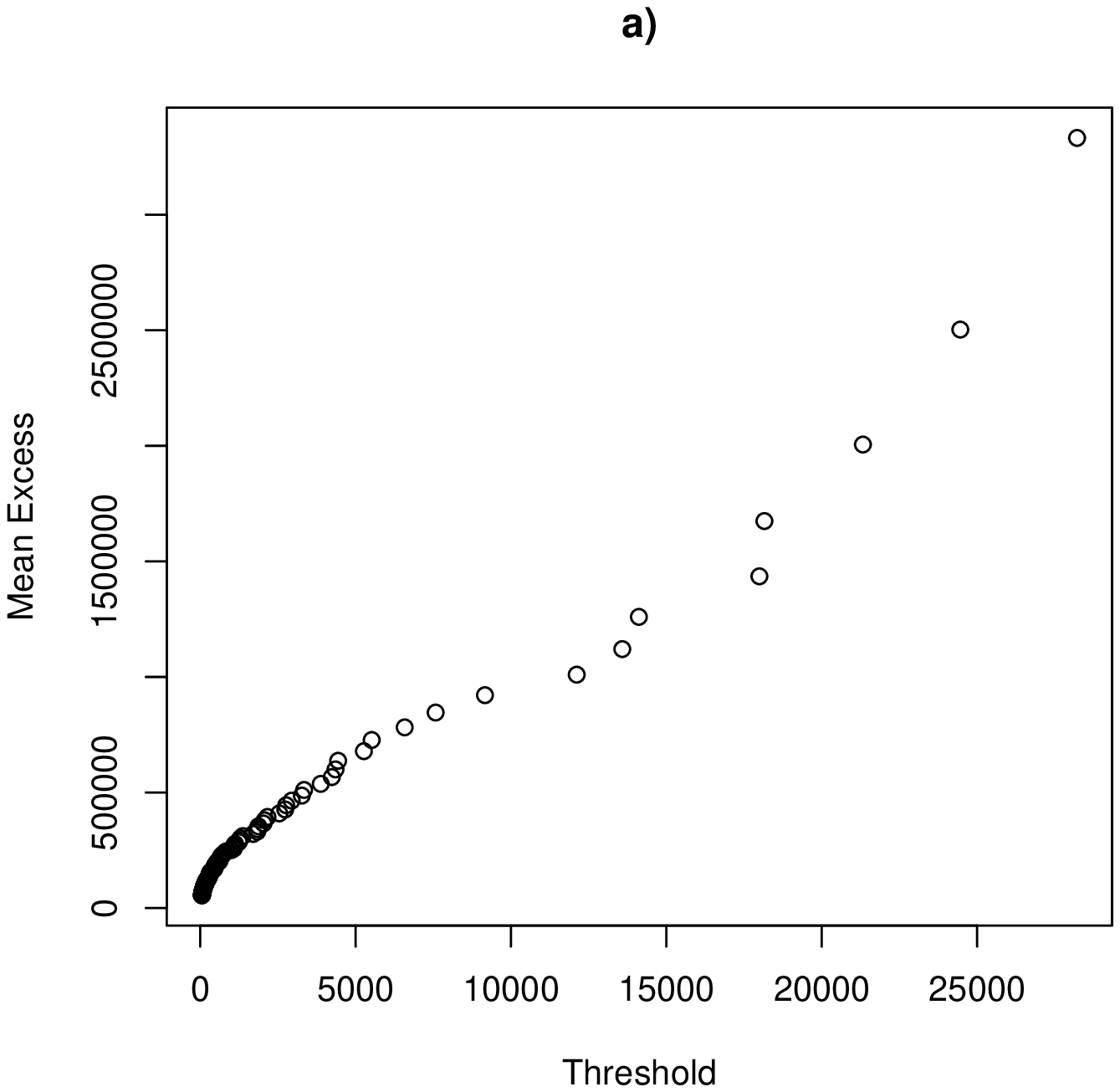}\vspace{-0.3cm}
\end{minipage}
\begin{minipage}[t]{0.45\linewidth}
    \includegraphics[scale=.3]{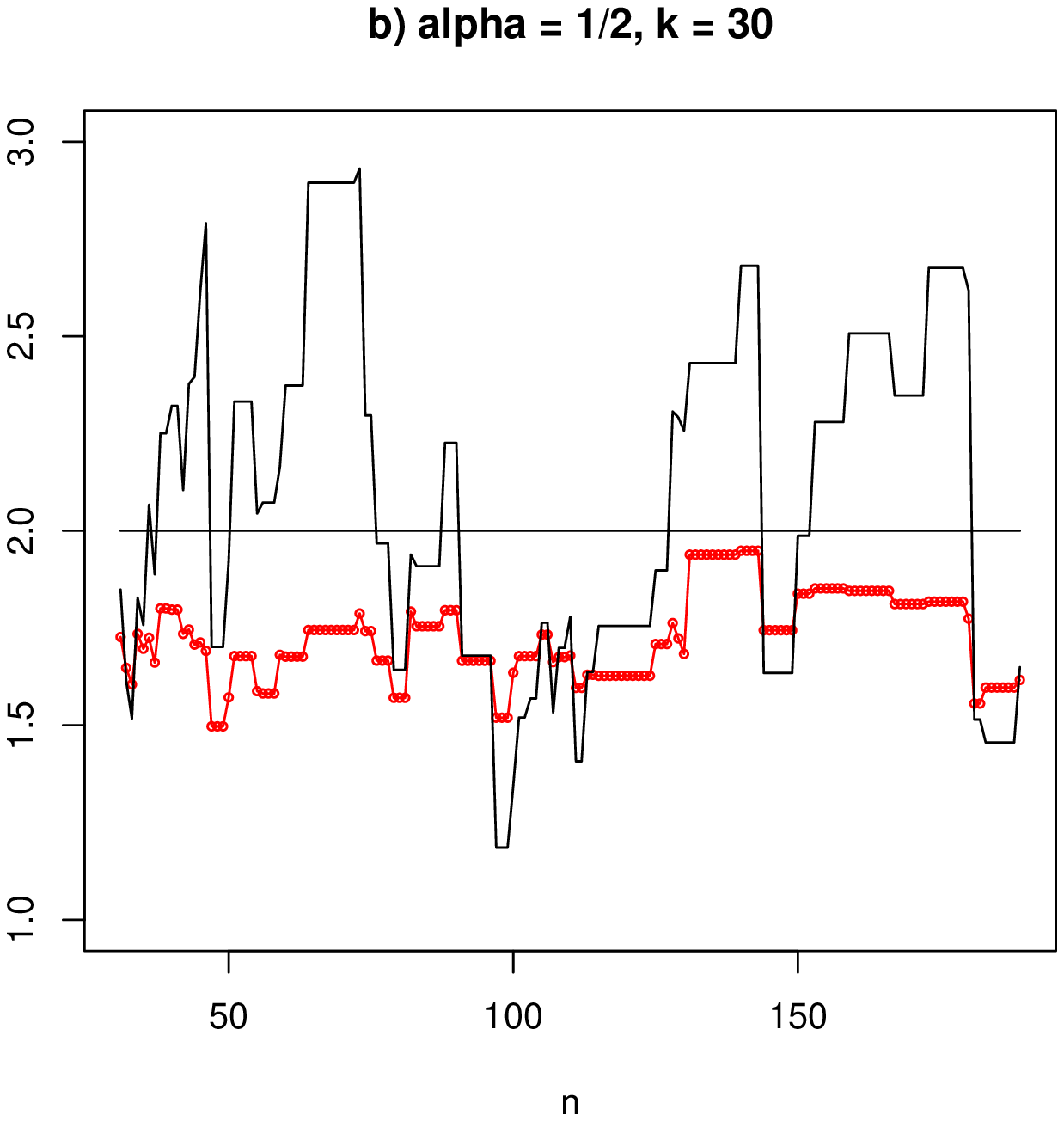}\vspace{-0.3cm}
\end{minipage}

\begin{minipage}[t]{0.45\linewidth}
   \includegraphics[scale=.3]{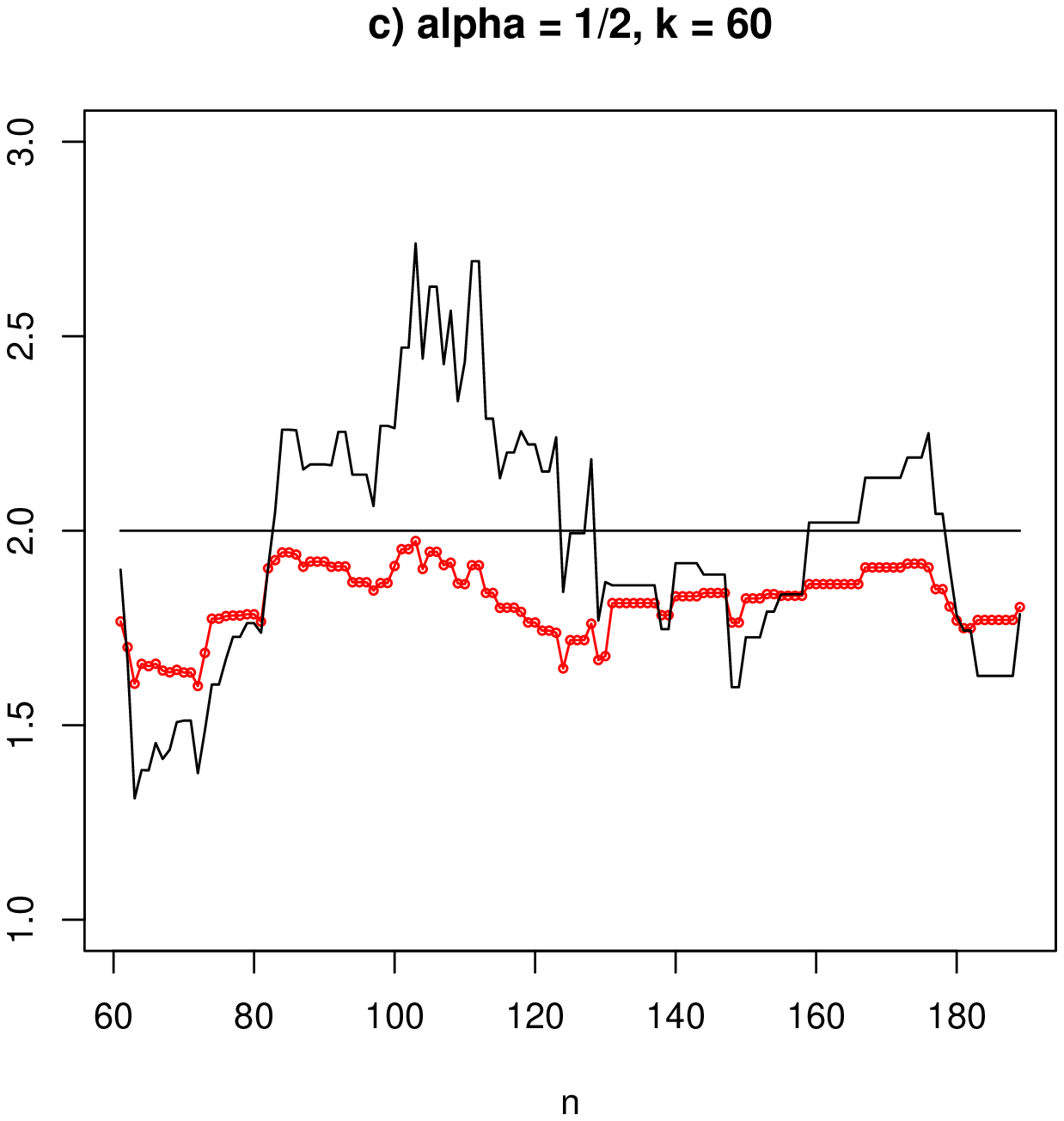}\vspace{-0.3cm}
\end{minipage}
\begin{minipage}[t]{0.45\linewidth}
    \includegraphics[scale=.3]{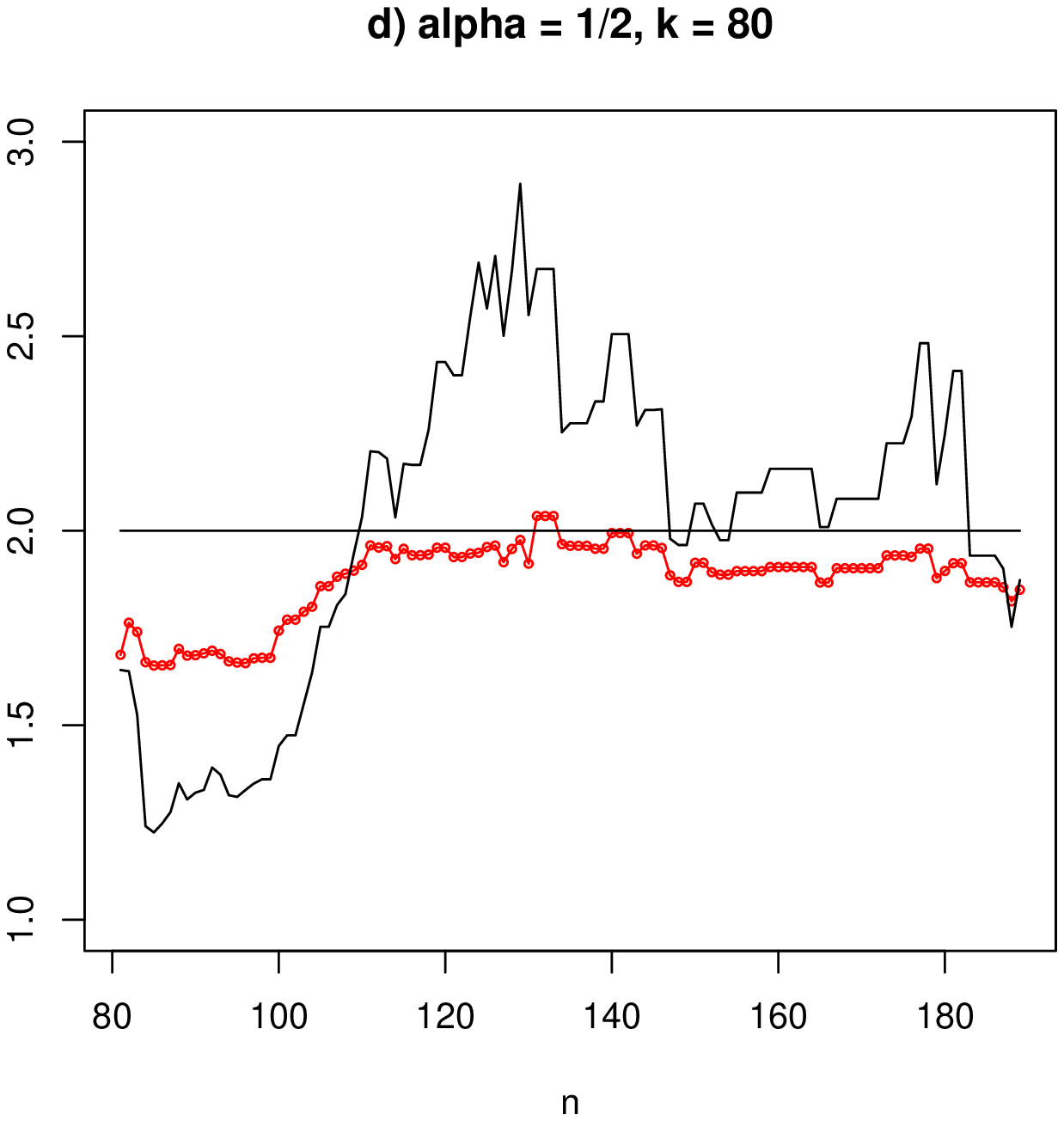}\vspace{-0.3cm}
\end{minipage}
\caption{Figures to Example 4, $\alpha = 0.5$}\label{fig:Figure1}
\end{center}
\end{figure}

\begin{ex}  \textbf{$\alpha = 0.5$.} Here we simulate $1500$ independent observations of a random variable with Hill horror distribution with parameter $\alpha = 0.5$. It is well known that the values of this distribution fluctuate too much and their expectation does not exist. Then we chose the threshold in such a way in order to have enough observations for CLT "to work". In this case we chose the threshold $50$ and obtain $n = 189$ observations above it. Having these observations we would like to estimate the tail of the distribution.  The plot of the mean excess function built on these $n = 189$  exceedances is given on Figure \ref{fig:Figure1} a).  Further on we plotted the Hill estimator (lines and dots) and Generalized Hill estimator for $p = -2$(lines) for fixed $k = 30$, $k = 60$ and $k = 80$ and different $n = k+1, ..., 189$. The straight line presents the true value of $\gamma = 2$.  The corresponding plots are given on Figures  \ref{fig:Figure1}, b),  c) and d).
\end{ex}

\begin{figure}
\begin{center}
\begin{minipage}[t]{0.45\linewidth}
    \includegraphics[scale=.3]{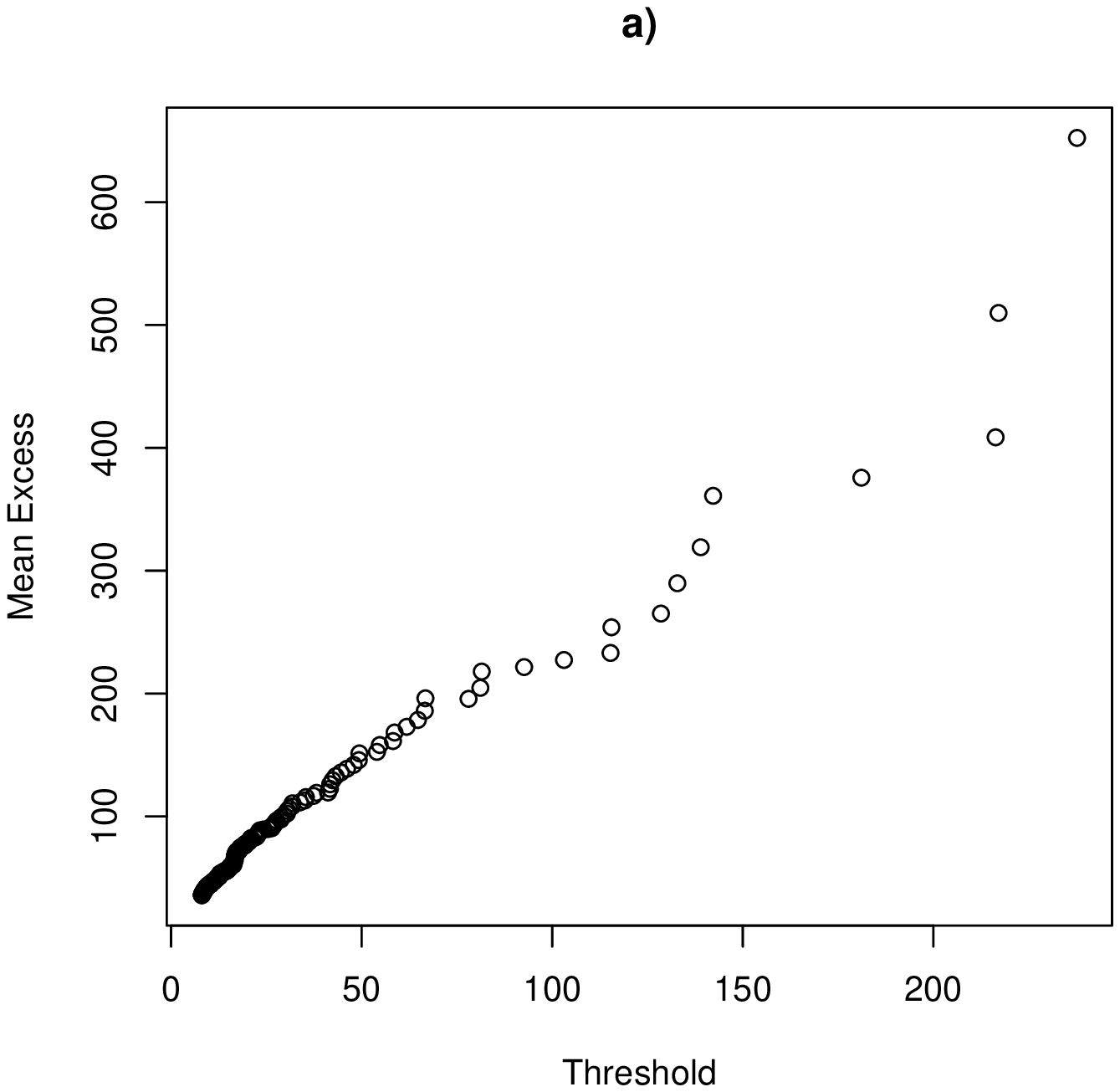}\vspace{-0.3cm}
\end{minipage}
\begin{minipage}[t]{0.45\linewidth}
    \includegraphics[scale=.3]{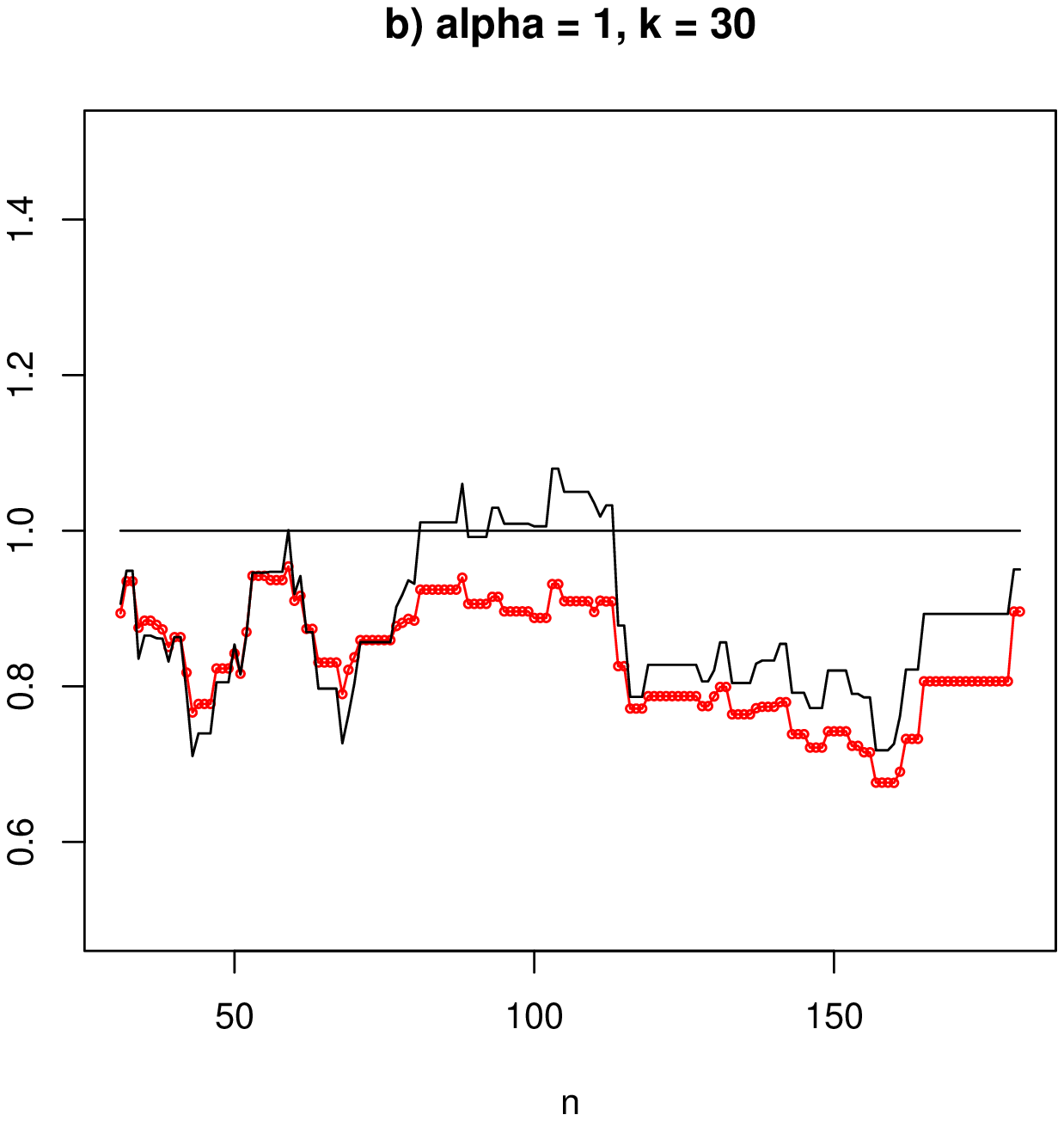}\vspace{-0.3cm}
\end{minipage}

\begin{minipage}[t]{0.45\linewidth}
   \includegraphics[scale=.3]{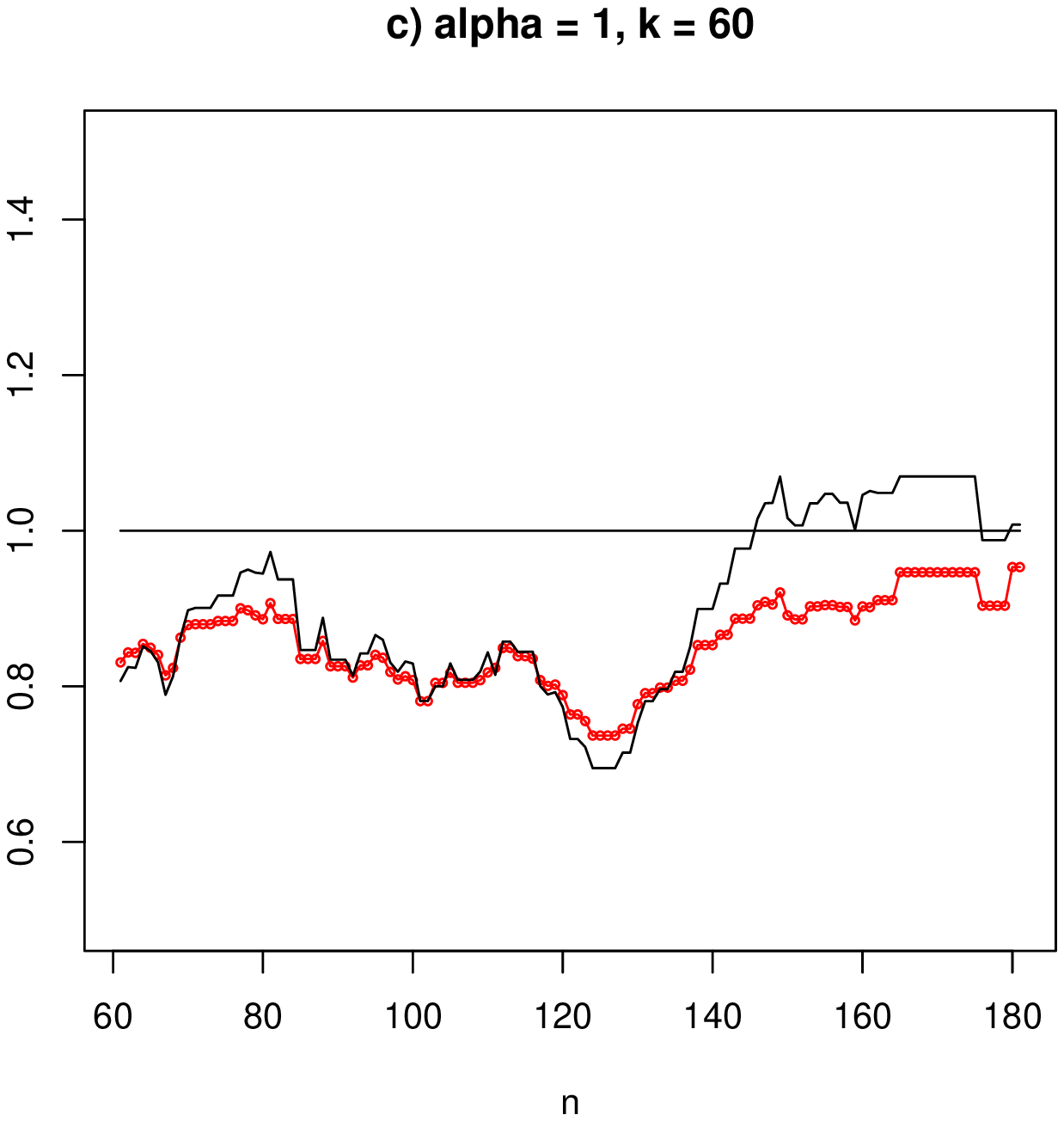}\vspace{-0.3cm}
\end{minipage}
\begin{minipage}[t]{0.45\linewidth}
    \includegraphics[scale=.3]{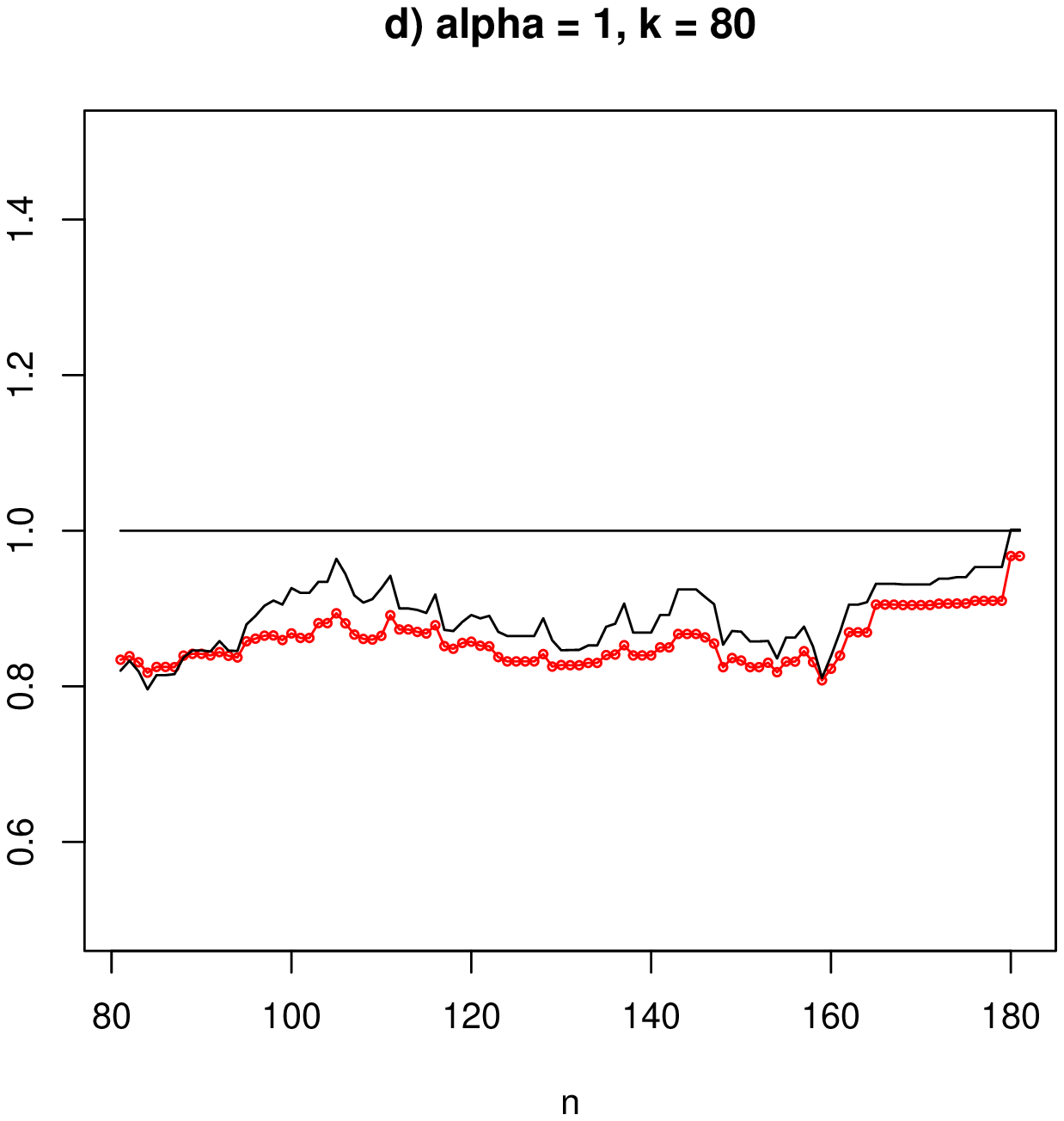}\vspace{-0.3cm}
\end{minipage}
\caption{Figures to Example {5, $\alpha = 1$}.}\label{fig:Figure2}
\end{center}
\end{figure}

\begin{ex}  \textbf{$\alpha = 1$.} {In this example we simulate $1500$ independent observations of  HillH(1) random variable. The values of this distribution fluctuate less than in the previous example, but the mean still does not exist. We chose the threshold {$10$} in such a way in order to have enough observations for CLT "to work" and obtain $n = 181$ observations above this threshold. Having these observations we would like to estimate the tail of the distribution.  The plot of the mean excess function built on these $n = 181$  exceedances is given on Figure \ref{fig:Figure2} a). The plots of the Hill estimator (lines and dots) and Generalized Hill estimator for $p = -1$(lines) for fixed $k = 30$, $k = 60$ and $k = 80$ and different $n = k+1, ..., 181$ are given on Figures  \ref{fig:Figure2}, b), c) and d). The straight line presents the true value of $\gamma = 1$. }
\end{ex}

\begin{figure}
\begin{center}
\begin{minipage}[t]{0.45\linewidth}
    \includegraphics[scale=.3]{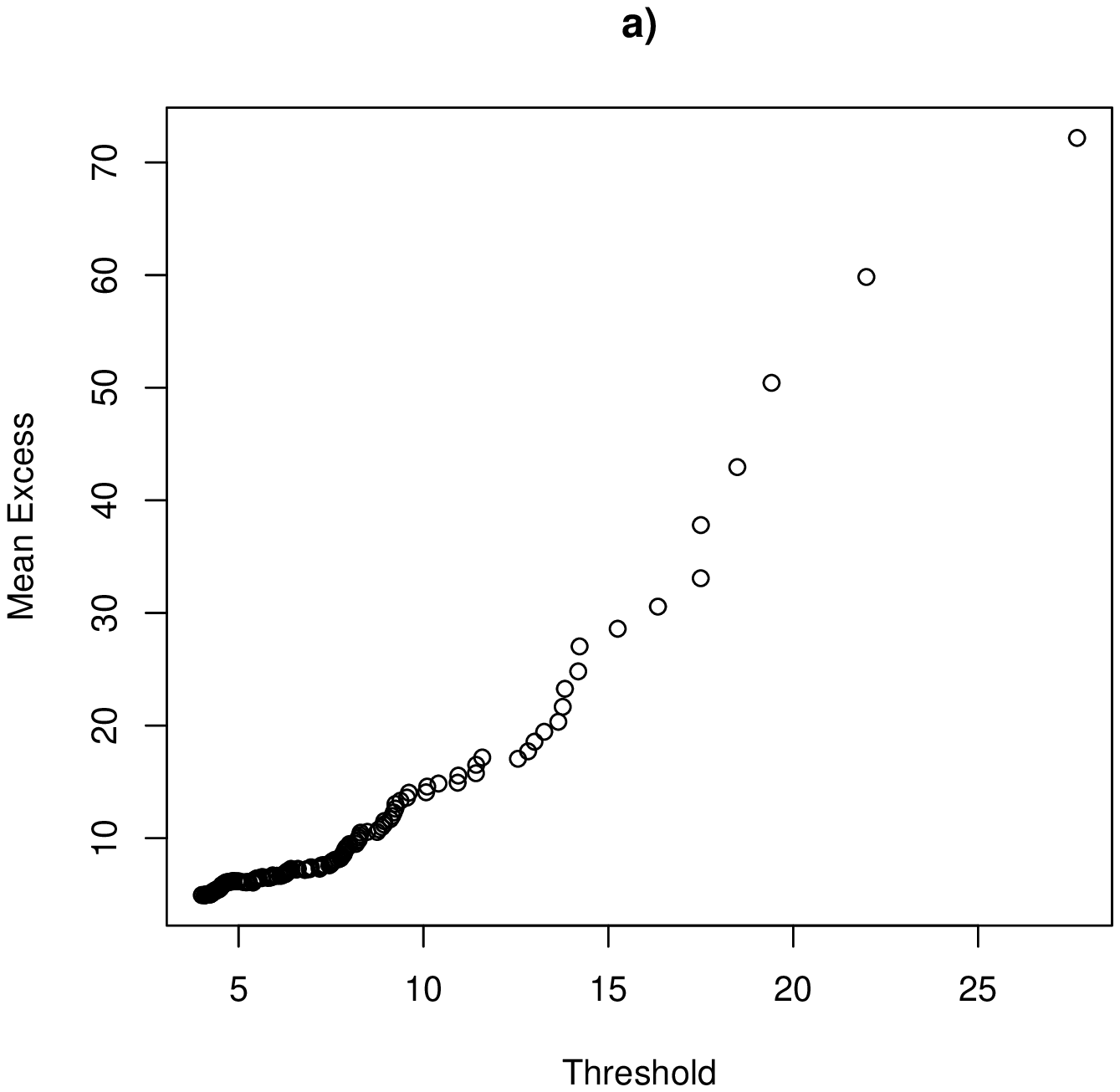}\vspace{-0.3cm}
\end{minipage}
\begin{minipage}[t]{0.45\linewidth}
    \includegraphics[scale=.3]{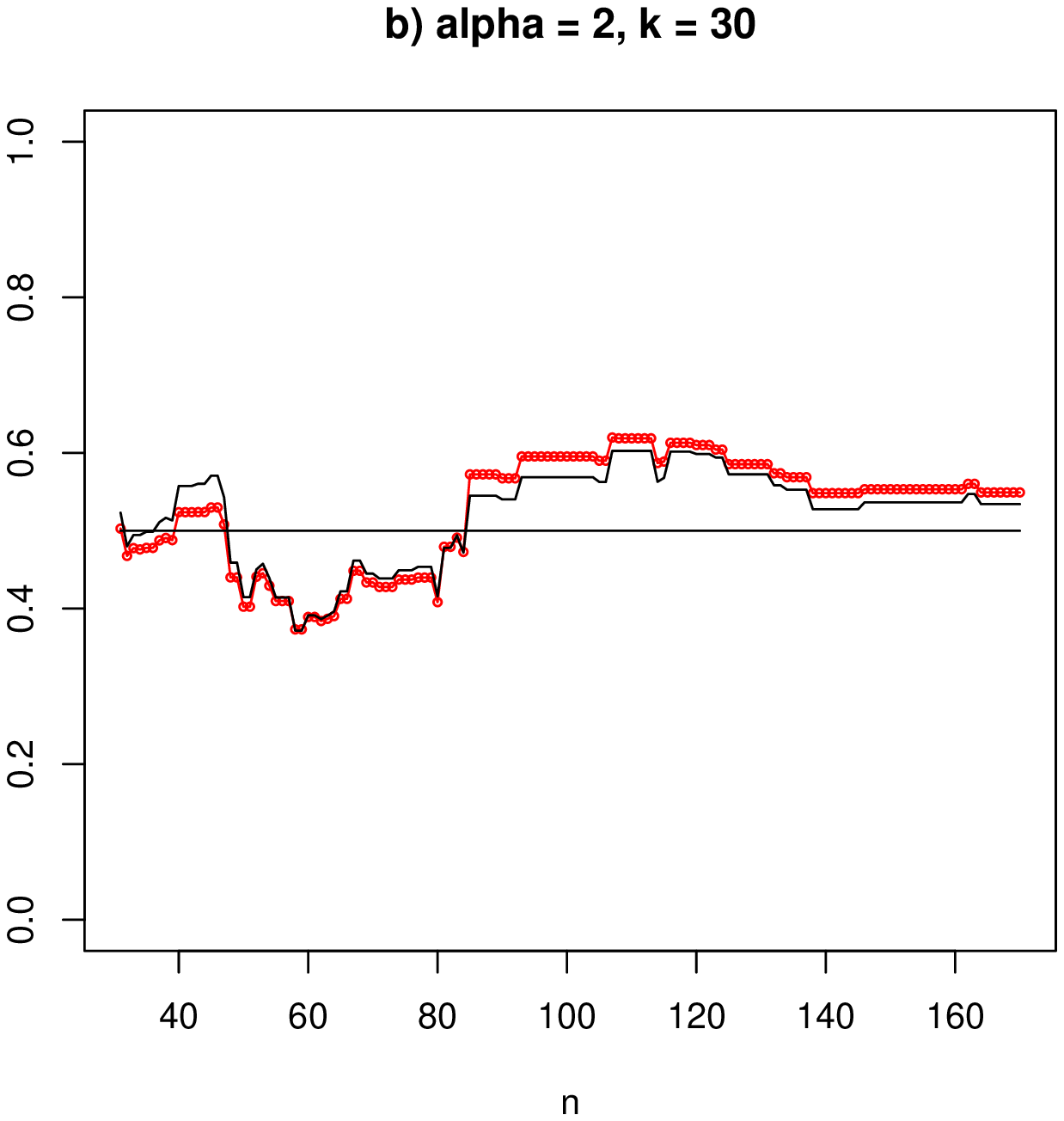}\vspace{-0.3cm}
\end{minipage}

\begin{minipage}[t]{0.45\linewidth}
   \includegraphics[scale=.3]{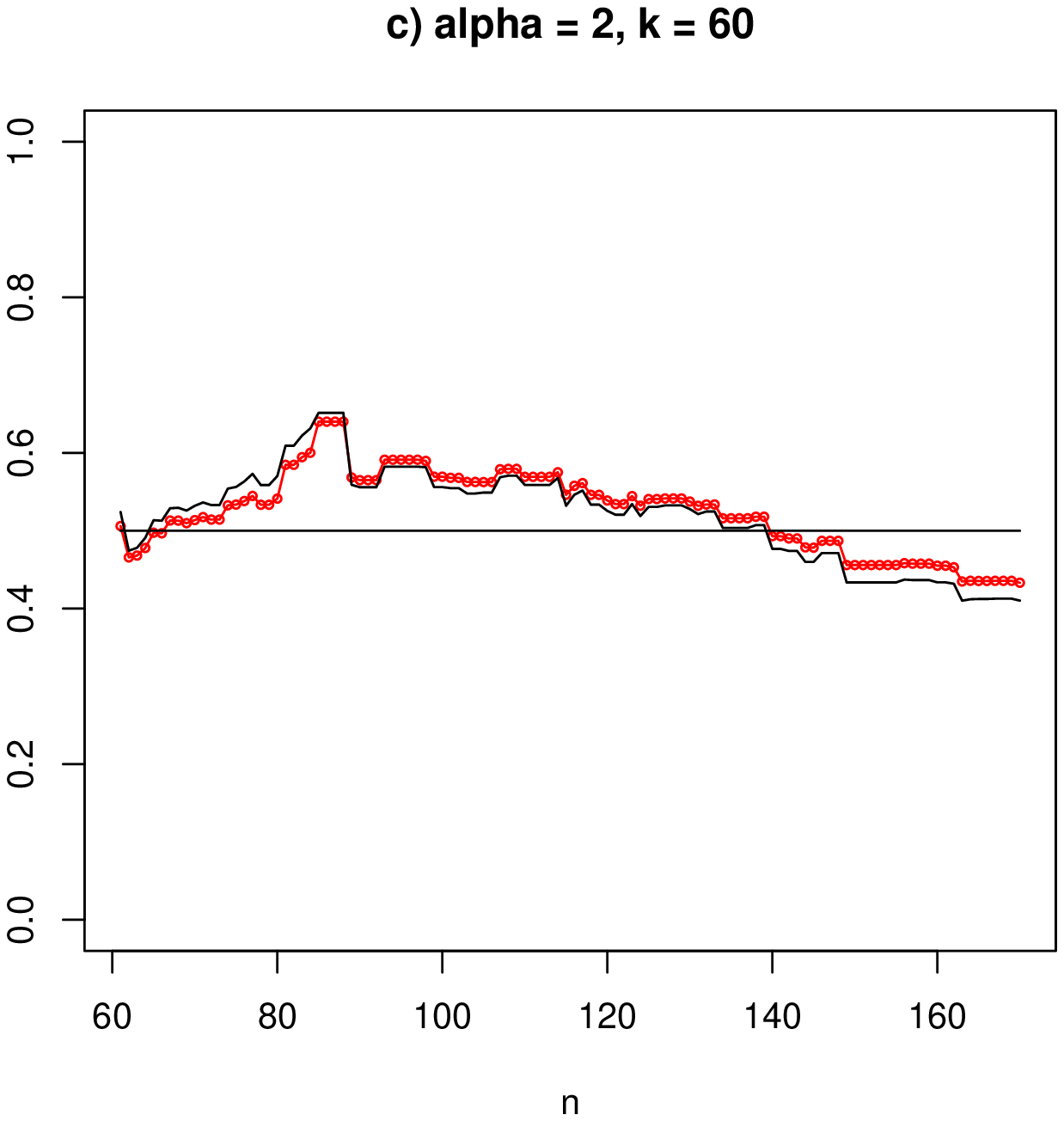}\vspace{-0.3cm}
\end{minipage}
\begin{minipage}[t]{0.45\linewidth}
    \includegraphics[scale=.3]{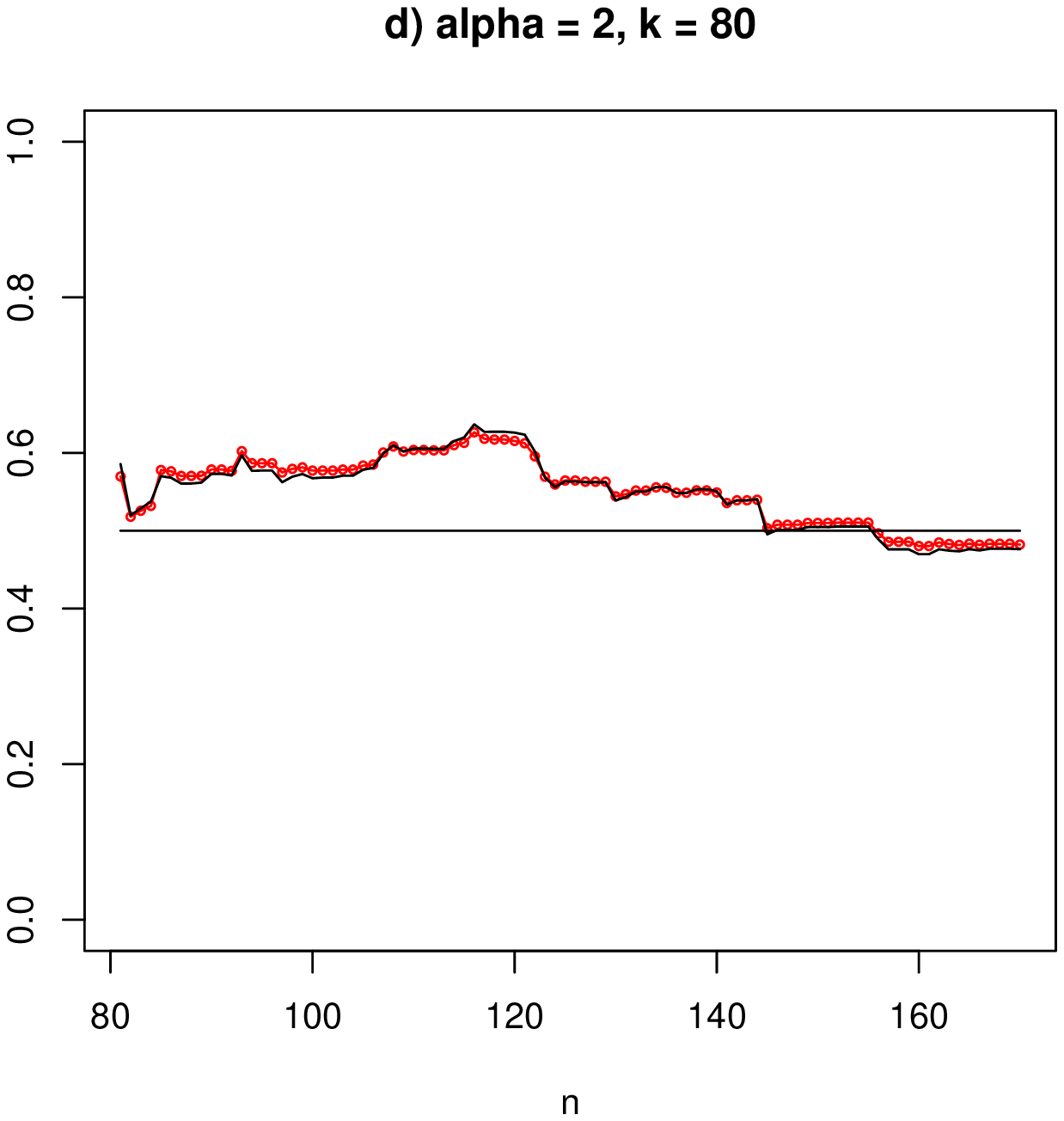}\vspace{-0.3cm}
    \end{minipage}

\caption{{ Figures to Example 6, $\alpha = 2$}.}\label{fig:Figure3}
\end{center}
\end{figure}

\begin{ex}  \textbf{$\alpha = 2$.} The variance of the  HillH(2) random variable presented in this example does not exist, but the expectation exists. We simulate $1500$ independent observations of HillH(2) random variable and again we chose appropriate threshold in order to have enough observations for CLT to give relatively good approximation. Here the threshold is $3$ and the number of the observations above it is $n = 170$.  Having these observations we would like to estimate $\gamma = 1/\alpha$. The plot of the mean excess function built on these $n = 170$  exceedances is given on Figure \ref{fig:Figure3}, a). The Hill estimator (lines and dots) and Generalized Hill estimator for $p = -0.5$(lines) for fixed $k = 30$, $k = 60$ and $k = 80$ and different $n = k+1, ..., 170$  are given on Figure \ref{fig:Figure3}, b), c) and d). The straight line again presents the true value of $\gamma = 0.5$.
\end{ex}

{\it Note:}  {In these examples} if we use the known form of the distribution, we do not need to chose any high threshold, in order to estimate $\gamma$ because the Hill horror distribution has Pareto tail. In that case we can also use the moment or other good estimators, however the situation is usually not such in practices. Therefore here we chose the "appropriate" threshold in order to follow the algorithm that is possible to apply to real data with unknown c.d.f.

\subsection*{Bootstrap techniques}

\begin{figure}
\begin{center}
\includegraphics[scale=.3]{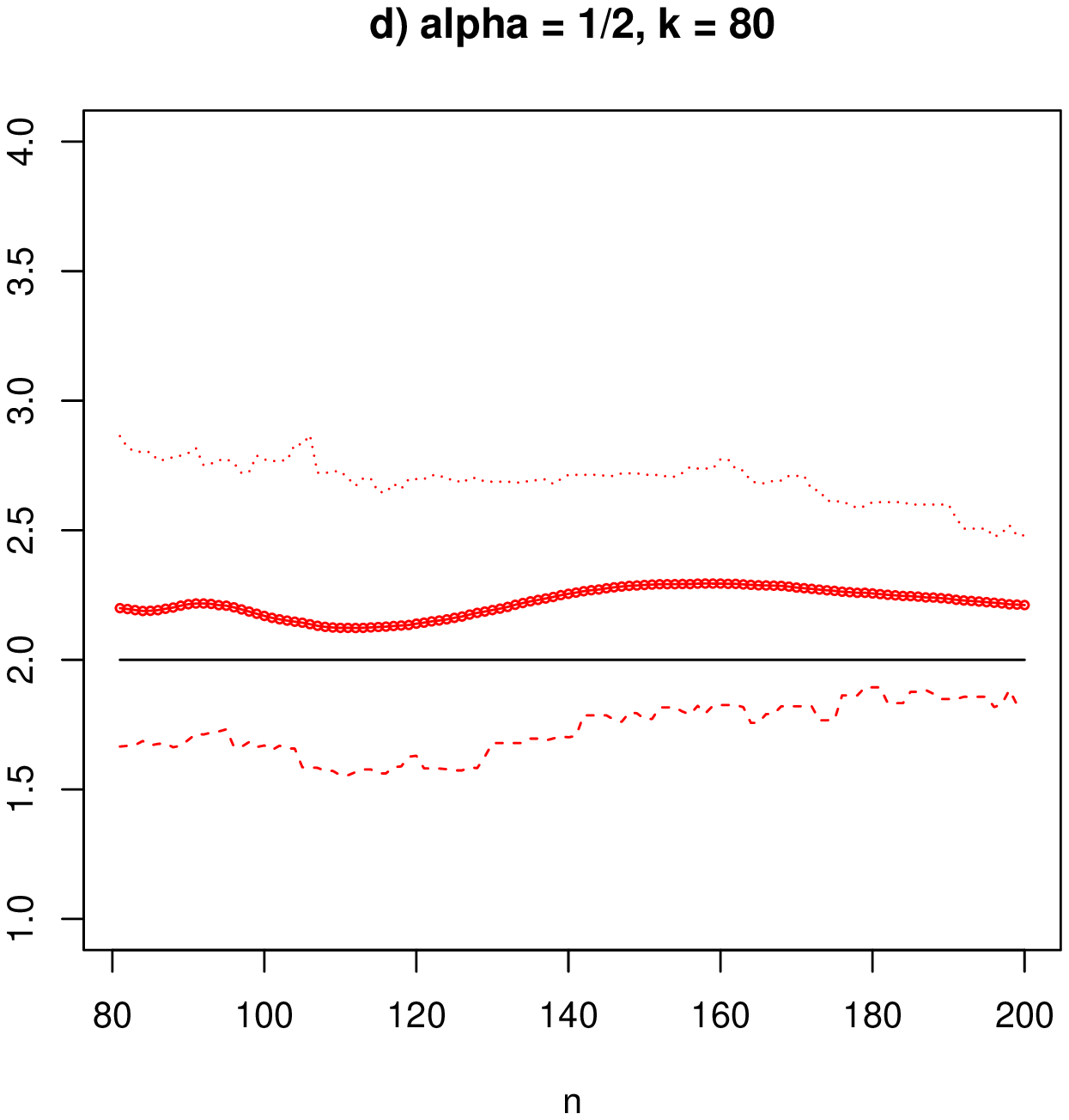} \includegraphics[scale=.3]{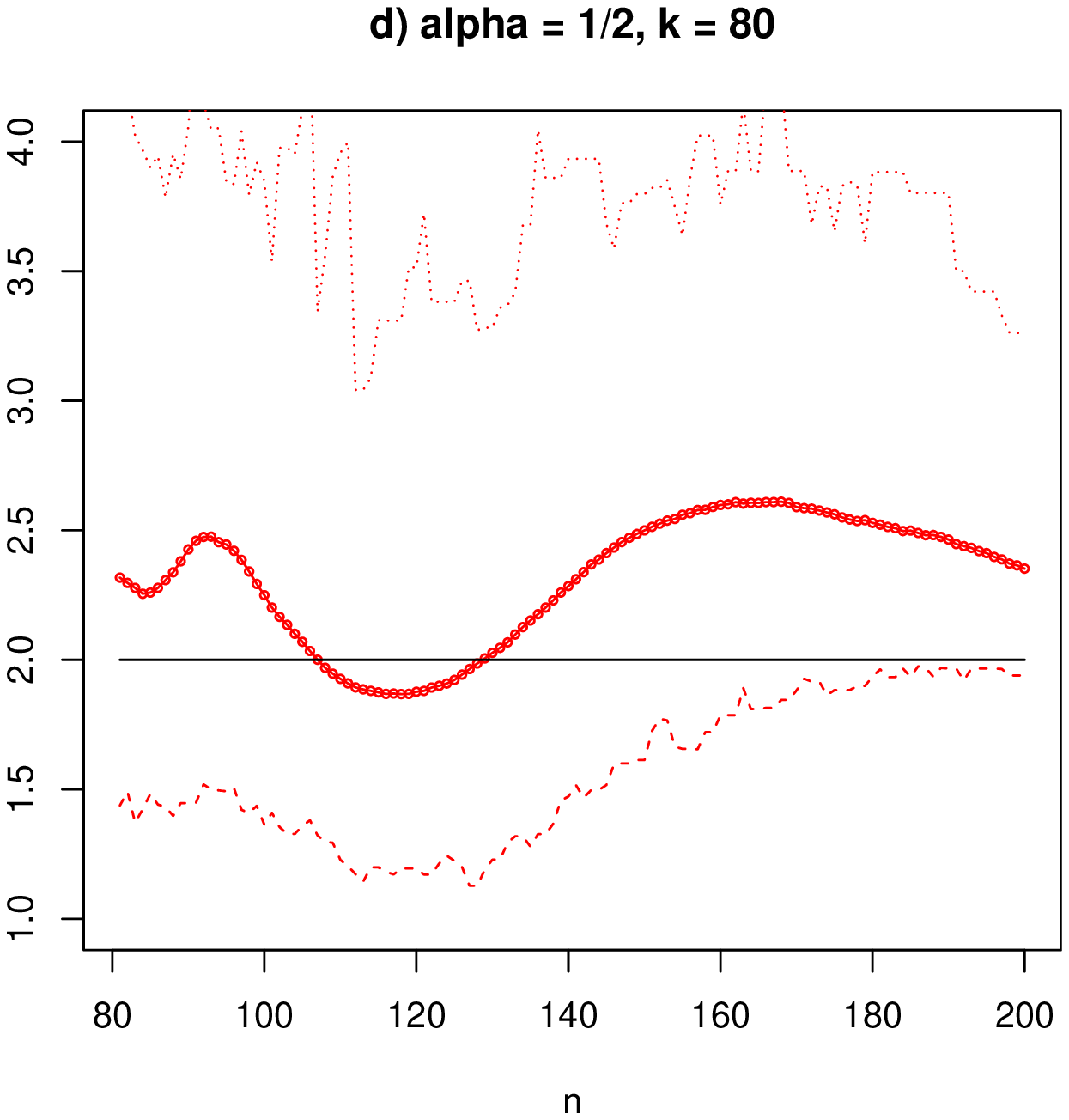}

\caption{$\alpha = 0.5$.The Hill estimator (left) and Generalized Hill estimator, for $p = -2$(right), fixed $k = 80$ and different $n$.  The straight line  presents the true value of $\gamma$.}\label{fig:Figure4}
\end{center}
\end{figure}

\begin{figure}
\begin{center}
\includegraphics[scale=.3]{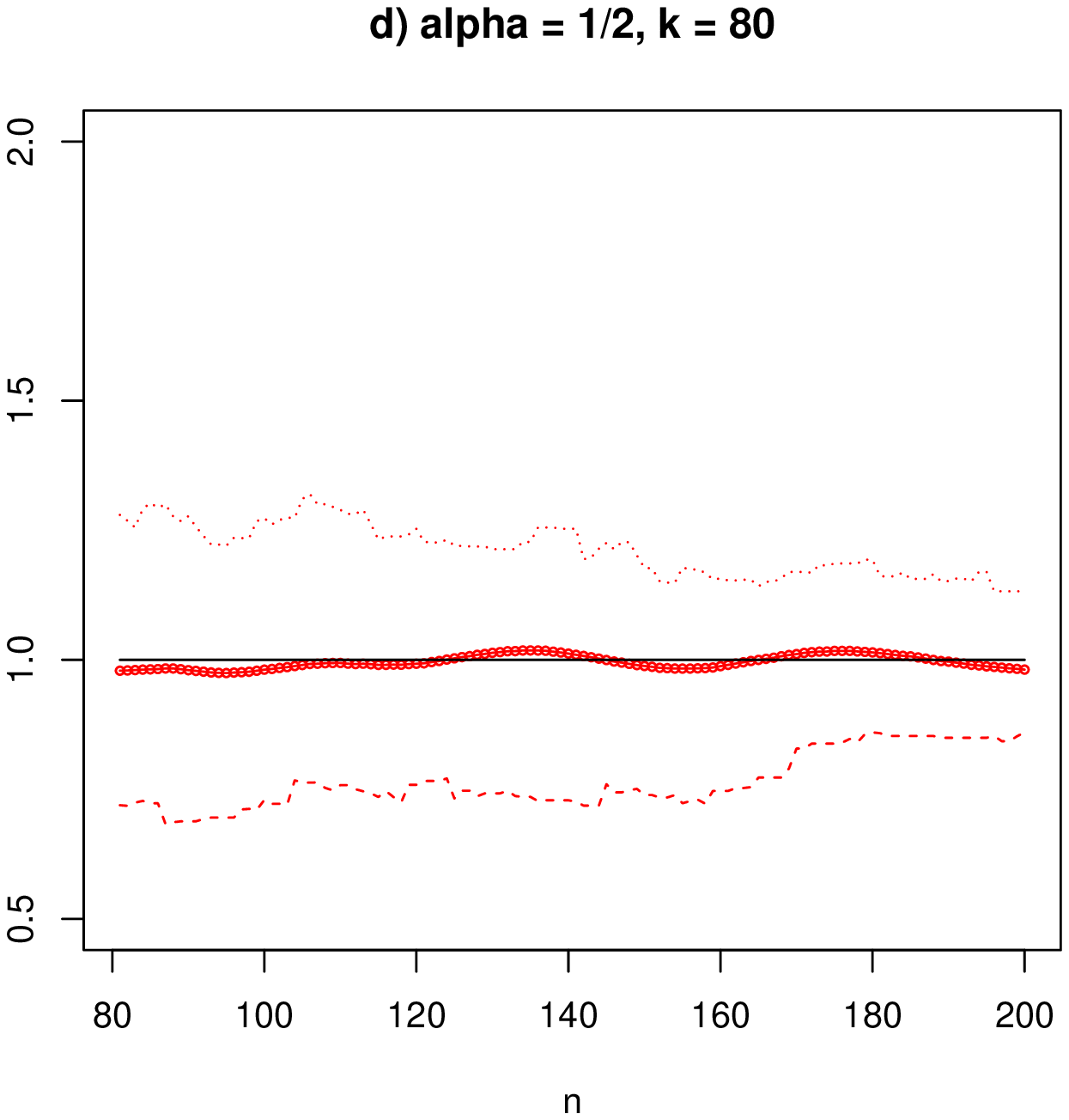} \includegraphics[scale=.3]{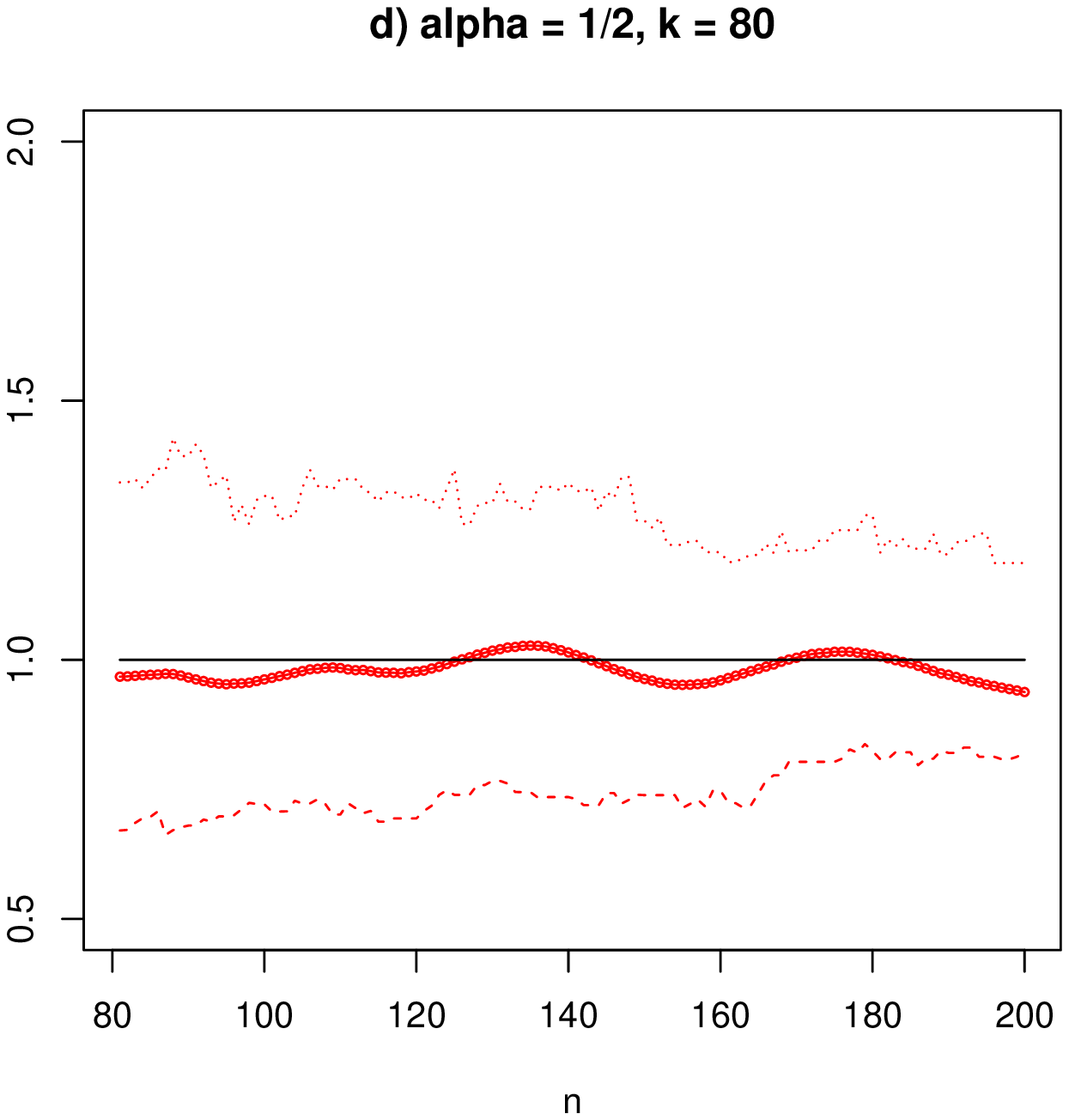}

\caption{$\alpha = 1$. The Hill estimator (left) and Generalised Hill estimator, for $p = -1$(right), fixed $k = 80$ and different $n$.  The straight line  presents the true value of $\gamma$.}\label{fig:Figure5}
\end{center}
\end{figure}

Having in mind the above considerations, in this subsection we make 1500 simulations of independent observations of the random variable with the Hill horror distribution with the corresponding parameter $\alpha$. Then we take 1000 different subsamples, without replacements and of sample size 1350  ($90\%$ of all observations), determine the threshold in such a way that to have $200$ its exceedances and for $k = 80$ and for $n = 81, 82, ..., 200$ we calculate Hill and the Generalized Hill estimators for these samples, $k$ and $n$. We repeat this procedure $1000$ times. Having these estimators for fixed $k$ and $n$ we calculate the averages of the Hill and the corresponding Generalized Hill estimators and plot them by - 0 - line, take their minima (- -) and maxima (dotted line) and again plot them. The resulting plots are given on Figures \ref{fig:Figure4}, \ref{fig:Figure5} and \ref{fig:Figure6}. The real estimated value $\gamma$ is given by straight line.

\begin{figure}[h]
\begin{center}
\begin{minipage}[t]{0.45\linewidth}
\includegraphics[scale=.3]{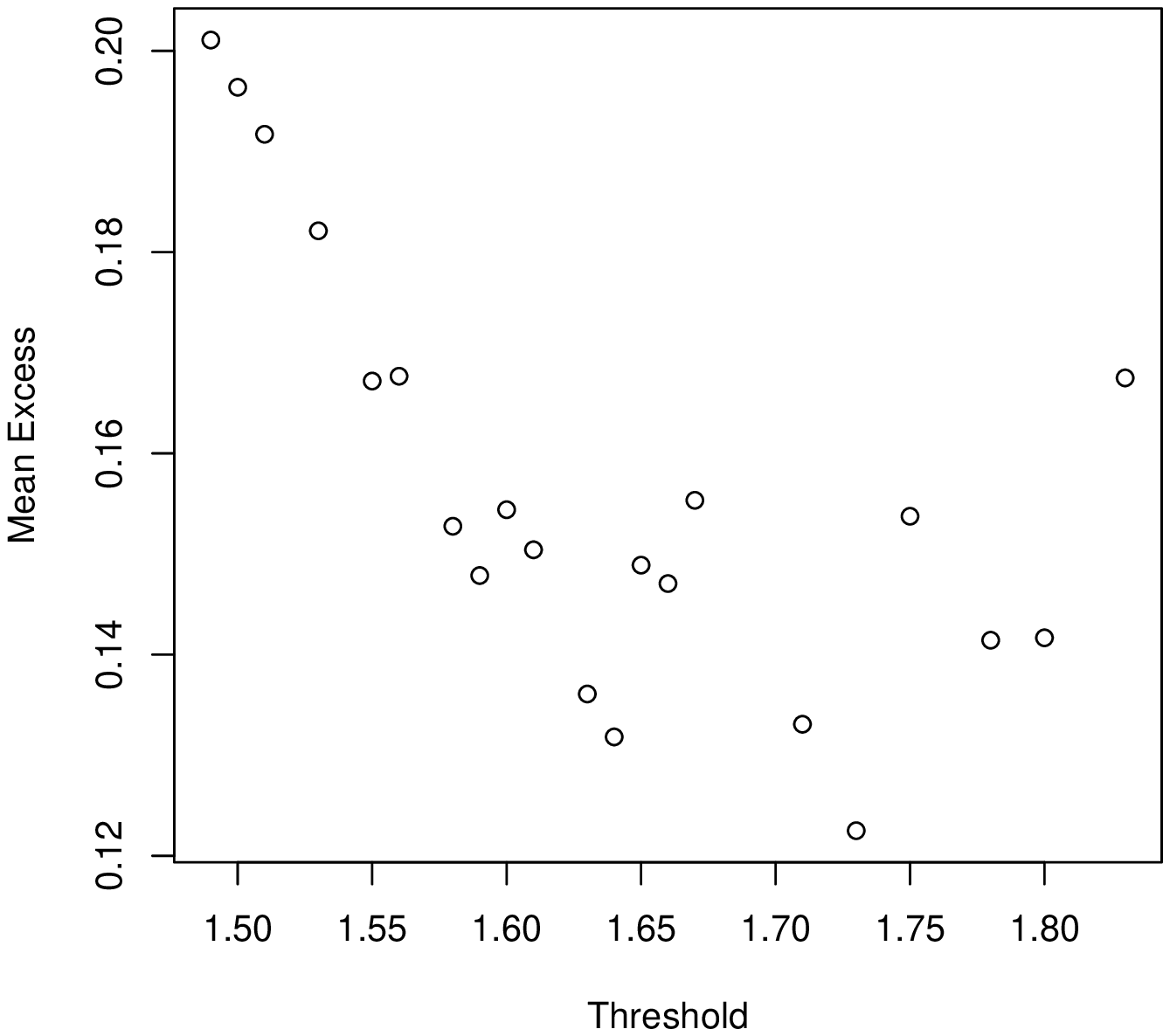}\vspace{-0.3cm}
\caption{Mean excess plot}
    \label{fig:MEanE}\end{minipage}
\begin{minipage}[t]{0.45\linewidth}
\includegraphics[scale=.3]{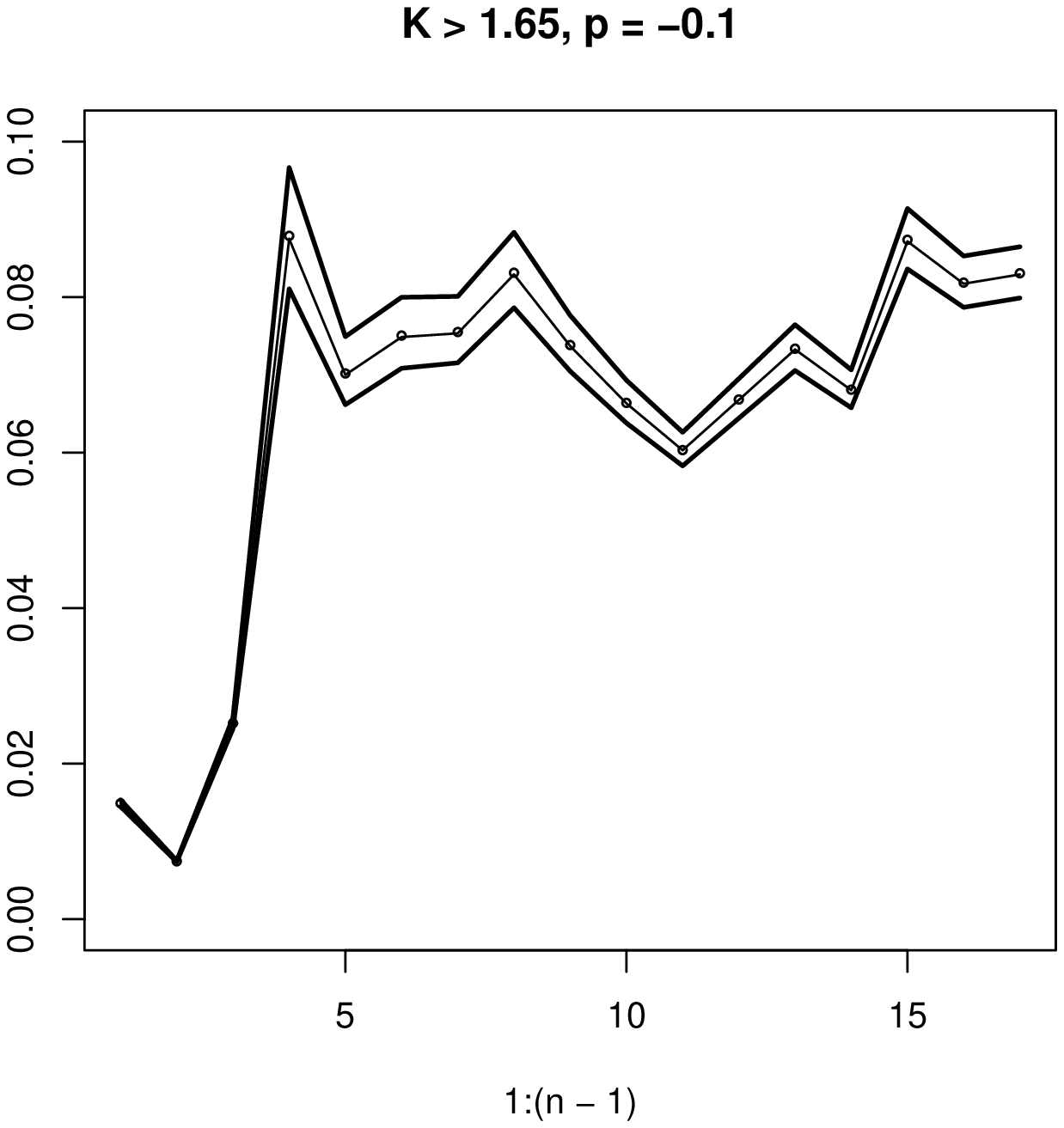}\vspace{-0.3cm}
\caption{Generalized Hill and Hill plots}
    \label{fig:GHP}
\end{minipage}
\end{center}
\end{figure}

\begin{figure}
\begin{center}
\includegraphics[scale=.3]{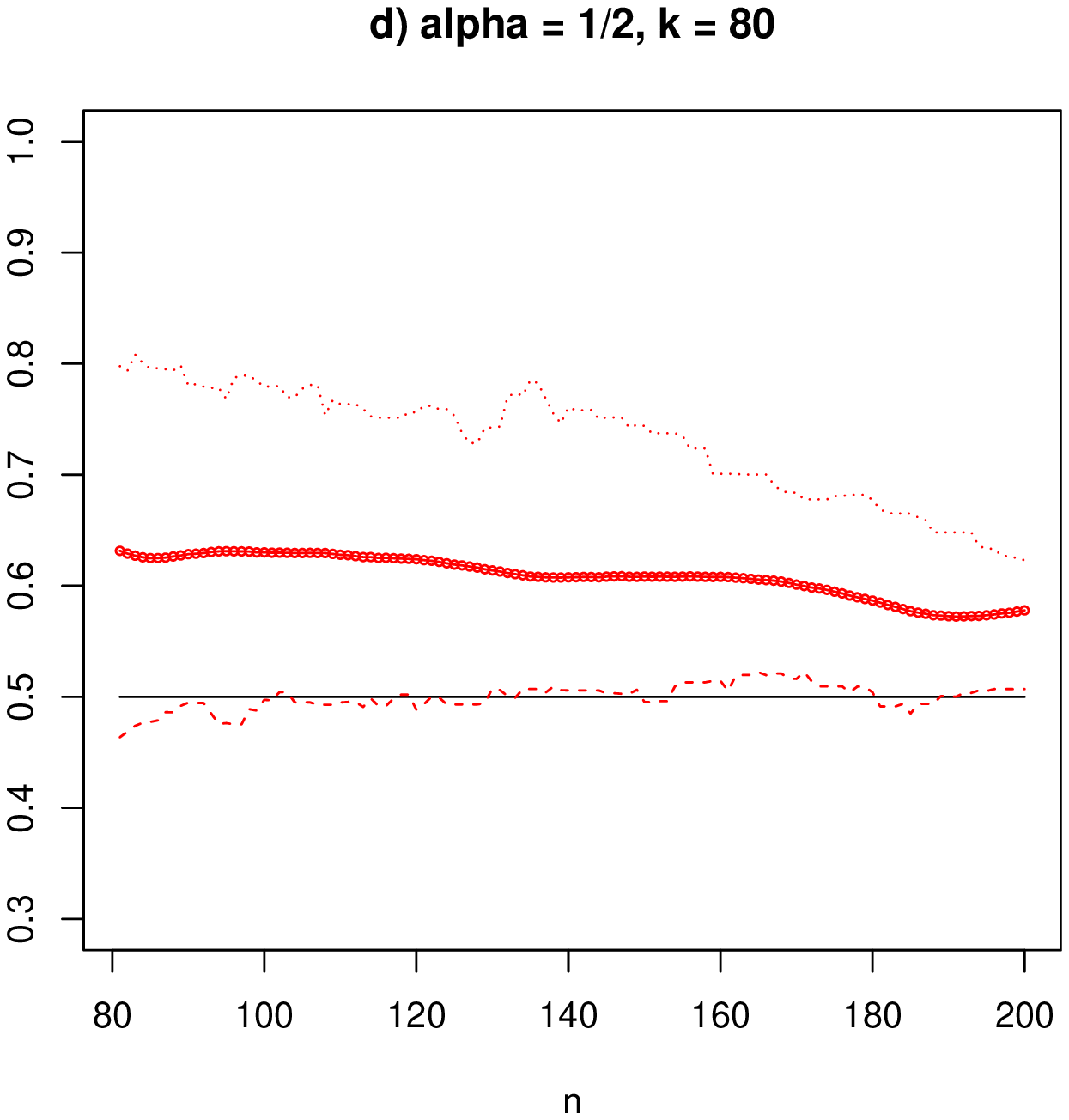} \includegraphics[scale=.3]{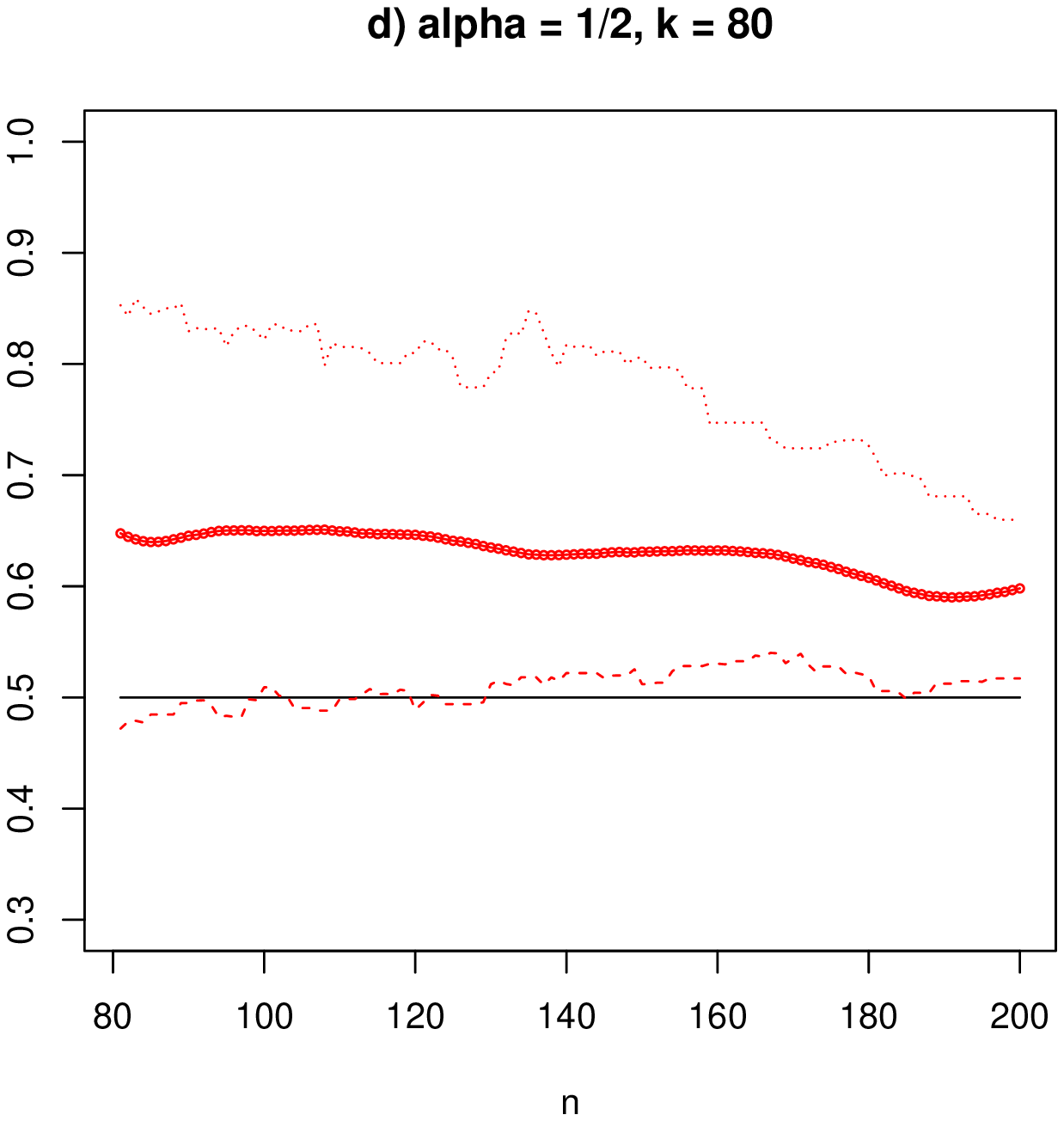}

\caption{ $\alpha = 2$, The Hill estimator (left) and Generalised Hill estimator, for $p = -0.5$(right), fixed $k = 80$ and different $n$.  The straight line  presents the true value of $\gamma$.}\label{fig:Figure6}
\end{center}
\end{figure}
Having this technique, using relatively small one initial sample of 1500 observations we obtain relatively good estimators of $\gamma$ and $\alpha$.

\section{Asymptotic normality of the Generalized Hill estimator}

The asymptotic normality of the Hill estimator is investigated by many authors. See e.g. \cite{HeauslerTeugels}. { They prove asymptotic normality of the Hill estimator without the second order regularly varying condition but the number of order statistics that participate in the estimator goes to $\infty$ in a very specific way. Most of the authors use the second order regularly varying condition in order to improve the rate of the convergence.} Here we consider the asymptotic normality of the generalized Hill estimator (the Hill estimator { could be considered as} a particular case for $p = 0$) without the second order regularly varying condition. In order to apply the Central Limit Theorem(CLT) let us remind the numerical characteristics of the limiting distributions in the previous theorem.
Let $U \sim U(0, 1)$, $\alpha > 0$, $\gamma = \frac{1}{\alpha}$ and $p \in R$, then
\begin{itemize}
\item[a)] For $s = 1, 2, ...$, $sp < \alpha$, i.e. $sp\gamma < 1$,
\begin{equation}\label{1}
E U^{-sp/\alpha} = \int_0^1 x^{-sp/\alpha} dx = \frac{x^{-sp/\alpha +1}}{-sp/\alpha +1}|_0^1 = \frac{\alpha}{\alpha - sp} = \frac{1}{1-sp\gamma}.
\end{equation}
\item[b)] For $\alpha \in R$ and $s = 1, 2, ...$,
\begin{equation}\label{2}
E \ln U^{-\frac{s}{\alpha}} = \frac{s}{\alpha} = s\gamma.
\end{equation}
\item[c)] For $2p < \alpha$, i.e. $p\gamma < \frac{1}{2}$,
\begin{equation}\label{3}
Var U^{-p/\alpha} = E U_i^{-2p/\alpha} - (E U_i^{-p/\alpha})^2 = \frac{1}{1-2p\gamma} - \frac{1}{(1-p\gamma)^2} = \frac{p^2\gamma^2}{(1-2p\gamma)(1-p\gamma)^2}.
\end{equation}
\item[d)] For $\alpha \in R$
\begin{equation}\label{4}
Var\,\, \ln U^{-\frac{1}{\alpha}} = \frac{1}{\alpha^2} = \gamma^2
\end{equation}
\end{itemize}

It is well known that the Pareto distribution does not satisfy the  second order regularly varying condition. In the section 2 we proved that the distribution of the Generalized Hill estimator in this case coincides with the one of the { transformed} average of $k$ i.i.d. Uniformly distributed random variables over the interval $(0, 1)$. Having in mind the CLT it is very natural to obtain that the distribution of the Generalized Hill estimator is asymptotically normal. In the following theorem we prove this result.

\begin{prop} Let $\mathbf{X}_1, \mathbf{X}_2, ...,
\mathbf{X}_n$ be independent copies of $\mathbf{X}$ with d.f. $F(x) = 1 -
x^{-\alpha}$, $x > 1$ and let $\Phi$ be the standard normal d.f. Then
\begin{enumerate}
\item[1.]
\begin{equation}\label{Pr51}
 \lim_{k \to \infty} \lim_{n \to \infty} P(\frac{\sqrt{k}(\hat{\gamma}_{X, k, n, 0} - \gamma)}{\gamma} < x)  = \Phi(x), \quad x \in \mathbb{R}.
\end{equation}
\item[2.]  for $2p < \alpha$, i.e. $p\gamma < \frac{1}{2}$,
\begin{equation}
 \lim_{k \to \infty} \lim_{n \to \infty}  P(\frac{\sqrt{k}(H_{X, k, n, p} - \frac{1}{1-p\gamma})}{\frac{p\gamma}{\sqrt{1-2p\gamma}(1-p\gamma)}} < x)  = \Phi(x), \quad x \in \mathbb{R}.
\end{equation}
\item[3.] for $2p < \alpha$, i.e. $p\gamma < \frac{1}{2}$,
{\begin{equation}\label{Pr52}
 \lim_{k \to \infty} \lim_{n \to \infty}  P(\frac{\sqrt{k}\left(\widehat{\gamma}_{X, k, n, p} - \gamma\right)}{\frac{\gamma(1-p\gamma)}{\sqrt{1-2p\gamma}}} < x)  = \Phi(x), \quad x \in \mathbb{R}.
\end{equation}
\item[4.] for $2p < \alpha$, i.e. $p\gamma < \frac{1}{2}$,
$$\sqrt{n-1}\frac{H_{X, n - 1, n, p} - \frac{1}{1-p\gamma }}{\frac{-p\gamma}{(-p\gamma + 1)\sqrt{-2\gamma p + 1}}} {\mathop{\to}\limits_{}^{d}} N(0, 1)$$
\begin{equation}\label{41} \lim_{n \to \infty}  P(\frac{\sqrt{n-1}\left(\widehat{\gamma}_{X, n-1, n, p} - \gamma\right)}{\frac{\gamma(1-p\gamma)}{\sqrt{1-2p\gamma}}} < x)  = \Phi(x), \quad x \in \mathbb{R}.\end{equation}
\begin{equation}\label{42}\lim_{n \to \infty} P(\frac{\sqrt{n-1}(\hat{\gamma}_{X, n-1, n, 0} - \gamma)}{\gamma} < x)  = \Phi(x), \quad x \in \mathbb{R}\end{equation}
}
\end{enumerate}
\end{prop}
{\bf Proof:} For the anyone of the following proofs we use Proposition 1 and the CLT{.}

1. By (\ref{2}) for $s = 1$ and (\ref{4}) we have
$$ \lim_{k \to \infty} \lim_{n \to \infty} P(\frac{\sqrt{k}(\hat{\gamma}_{X, k, n, 0} - \gamma)}{\gamma} < x) = \lim_{k \to \infty} P(\frac{\sqrt{k}(\frac{1}{k}\sum_{i=1}^{k} \ln \,U_i^{-\frac{1}{\alpha }} - \gamma)}{\gamma} < x) = \Phi(x), \quad x \in \mathbb{R}.$$

2. By (\ref{1}) for $s = 1$ and (\ref{3}) we have
$$\lim_{k \to \infty} \lim_{n \to \infty}  P(\frac{\sqrt{k}(H_{X, k, n, p} - \frac{1}{1-p\gamma})}{\frac{p\gamma}{\sqrt{1-2p\gamma}(1-p\gamma)}} < x) = \lim_{k \to \infty}  P(\frac{\sqrt{k}(\frac{1}{k}\sum_{i=1}^{k}U_i^{-\frac{p}{\alpha }} - \frac{1}{1-p\gamma})}{\frac{p\gamma}{\sqrt{1-2p\gamma}(1-p\gamma)}} < x) = \Phi(x), \quad x \in \mathbb{R}.$$

3. Here we consider the function $h(y) = \frac{1}{p}\left(1-\frac{1}{y}\right)$ with derivative $h'(y) = \frac{1}{py^2}$, apply the delta method and obtain
{ $$\lim_{k \to \infty} P(\left(\frac{\sqrt{k}\left(\widehat{\gamma}_{X, k, n, p} - \gamma\right)}{\sqrt{\frac{1}{p^2\left(\frac{1}{1 - p\gamma}\right)^4}\frac{p^2\gamma^2}{(1-2p\gamma)(1-p\gamma)^2}}} < x \right)\,  = \lim_{k \to \infty} P(\left(\frac{\sqrt{k}\left(\frac{1}{p}\left(1 - \frac{1}{H_{X, k, n, p}}\right) - \gamma\right)}{\sqrt{\frac{1}{p^2\left(\frac{1}{1 - p\gamma}\right)^4}\frac{p^2\gamma^2}{(1-2p\gamma)(1-p\gamma)^2}}} < x \right)\, =$$ }
$$ = \, \lim_{k \to \infty} P\left(\frac{\sqrt{k}\left(\frac{1}{p}\left(1 - \frac{1}{H_{X, k, n, p}}\right) - \gamma\right)}
{\frac{\gamma(1-p\gamma)}{\sqrt{1-2p\gamma}}} < x \right) = \Phi(x), \quad x \in \mathbb{R}.$$

{
4. According to Proposition 1
$$ H_{X, n - 1, n, p} \,{\mathop{=}\limits_{}^{d}}\, \frac{1}{n - 1}\sum_{i=1}^{n - 1}U_i^{-p\gamma}.$$
Let us now apply the CLTh
$$\sqrt{n-1}\frac{H_{X, n - 1, n, p} - \frac{1}{1-p\gamma}}{\frac{-p\gamma}{(1 - p\gamma)\sqrt{1 - 2p\gamma}}} {\mathop{\to}\limits_{}^{d}} N(0, 1)$$
We apply the delta method and obtain the desired result.

In analogous way we prove the corresponding statement for the Hill estimator.
}

 \hfill $\Box$

{\it Note:} 1. Now we obtain the following confidence intervals for $\gamma$
 having  $k$ and $n$ large enough such that  $k < < n$,
\begin{equation}\label{conf_int}
\left(\frac{\hat{\gamma}_{X, k, n, 0}}{\frac{z_{1 - \alpha/2}}{\sqrt{k}} + 1}; \frac{\hat{\gamma}_{X, k, n, 0}}{\frac{z_{\alpha/2}}{\sqrt{k}} + 1}\right)
\end{equation}

{2. The statement 4. shows that it is not obligatory $k$ to be infinitely small function of $n$ in order to obtain asymptotic normality without random centering and without second order regularly varying condition.

3. Note that (\ref{Pr51}) is just a particular case of (\ref{Pr52}) and (\ref{42}) is a particular case of (\ref{41}) for $p = 0$.}

Further on we generalize these results for any distribution with regularly varying tail.

\begin{thm} Let $\mathbf{X}_1, \mathbf{X}_2, ...,
\mathbf{X}_n$ be independent copies of $\mathbf{X}$ with d.f. $\overline{F} \in RV_{-\alpha}$,  and let $\Phi$ be the standard normal d.f. Then
\begin{enumerate}
\item[1.]
\begin{equation}
 \lim_{k \to \infty} \lim_{n \to \infty}  P(\frac{\sqrt{k}(\hat{\gamma}_{X, k, n, 0} -\gamma)}{\gamma} < x)  = \Phi(x), \quad x \in \mathbb{R}.
\end{equation}
\item[2.]  $2p < \alpha$, i.e. $p\gamma < \frac{1}{2}$,
\begin{equation}
 \lim_{k \to \infty} \lim_{n \to \infty}  P(\frac{\sqrt{k}(H_{X, k, n, p} - \frac{1}{1-p\gamma})}{\frac{p\gamma}{\sqrt{1-2p\gamma}(1-p\gamma)}} < x)  = \Phi(x), \quad x \in \mathbb{R}.
\end{equation}
\item[3.]  $2p < \alpha$, i.e. $p\gamma < \frac{1}{2}$,
{\begin{equation}
 \lim_{k \to \infty} \lim_{n \to \infty}  P(\frac{\sqrt{k}\left(\widehat{\gamma}_{X, k, n, p} - \gamma\right)}{\frac{\gamma(1-p\gamma)}{\sqrt{1-2p\gamma}}} < x)  = \Phi(x), \quad x \in \mathbb{R}.
\end{equation}}
\end{enumerate}
\end{thm}

{\bf Proof:} For the anyone of the following proofs we use Proposition 2 and the CLT

1. By (\ref{2}) for $s = 1$ and (\ref{4}) we have
$$ \lim_{k \to \infty} \lim_{n \to \infty} P(\frac{\sqrt{k}(\hat{\gamma}_{X, k, n, 0} - \gamma)}{\gamma} < x) =
\lim_{k \to \infty} \lim_{n \to \infty} P(\hat{\gamma}_{X, k, n, 0} < x\frac{\gamma}{\sqrt{k}} + \gamma) =$$
$$= \lim_{k \to \infty} P(\frac{1}{k}\sum_{i=1}^{k} \ln \,U_i^{-\frac{1}{\alpha }} < x\frac{\gamma}{\sqrt{k}} + \gamma) = \lim_{k \to \infty} P(\frac{\sqrt{k}(\frac{1}{k}\sum_{i=1}^{k} \ln \,U_i^{-\frac{1}{\alpha }} - \gamma)}{\gamma} < x) = \Phi(x), \quad x \in \mathbb{R}.$$

2. By (\ref{1}) for $s = 1$ and (\ref{3}) we have
$$\lim_{k \to \infty} \lim_{n \to \infty}  P(\frac{\sqrt{k}(H_{X, k, n, p} - \frac{1}{1-p\gamma})}{\frac{p\gamma}{\sqrt{1-2p\gamma}(1-p\gamma)}} < x) = $$
$$ = \lim_{k \to \infty} \lim_{n \to \infty}  P(H_{X, k, n, p} < x\frac{p\gamma}{\sqrt{k}\sqrt{1-2p\gamma}(1-p\gamma)} +  \frac{1}{1-p\gamma}) = $$
$$ = \lim_{k \to \infty} P(\frac{1}{k}\sum_{i=1}^{k}U_i^{-\frac{p}{\alpha }} < x\frac{p\gamma}{\sqrt{k}\sqrt{1-2p\gamma}(1-p\gamma)} +  \frac{1}{1-p\gamma}) =$$
$$ = \lim_{k \to \infty}  P(\frac{\sqrt{k}(\frac{1}{k}\sum_{i=1}^{k}U_i^{-\frac{p}{\alpha }} - \frac{1}{1-p\gamma})}{\frac{p\gamma}{\sqrt{1-2p\gamma}(1-p\gamma)}} < x) = \Phi(x), \quad x \in \mathbb{R}.$$

3. Here we consider the function $h(y) = \frac{1}{p}\left(1-\frac{1}{y}\right)$ with derivative $h'(y) = \frac{1}{py^2}$, apply the delta method and obtain
{ $$\lim_{k \to \infty} P(\left(\frac{\sqrt{k}\left(\widehat{\gamma}_{X, k, n, p} - \gamma\right)}{\sqrt{\frac{1}{p^2\left(\frac{1}{1 - p\gamma}\right)^4}\frac{p^2\gamma^2}{(1-2p\gamma)(1-p\gamma)^2}}} < x \right)\,  = \lim_{k \to \infty} P(\left(\frac{\sqrt{k}\left(\frac{1}{p}\left(1 - \frac{1}{H_{X, k, n, p}}\right) - \gamma\right)}{\sqrt{\frac{1}{p^2\left(\frac{1}{1 - p\gamma}\right)^4}\frac{p^2\gamma^2}{(1-2p\gamma)(1-p\gamma)^2}}} < x \right)\, =$$ }
$$ =\, \lim_{k \to \infty} P\left( \frac{\sqrt{k}\left(\frac{1}{p}\left(1 - \frac{1}{H_{X, k, n, p}}\right) - \gamma\right)}
{\frac{\gamma(1-p\gamma)}{\sqrt{1-2p\gamma}}} < x\right) = \Phi(x), \quad x \in \mathbb{R}.$$

 \hfill $\Box$

{ {\it Note:}  Again 1. is just a particular case of 3. for $p = 0$.}

\section{How to find the most appropriate $p$?}

In this section we discuss how to find the most appropriate $p$ for the generalized Hill estimator.

If we consider the smallest asymptotic variance of the Generalized Hill estimator, using Theorem 2., 3) we obtain that if $p$ goes to $0$ the  variance goes to its minima which is 1. More precisely if we consider the function
$$f(p) = \frac{\gamma(1-p\gamma)}{\sqrt{1-2p\gamma}},$$
for $p < \frac{1}{2\gamma} = \frac{\alpha}{2}$ then
$$f'(p) = \frac{\gamma^3 p}{\sqrt{(1-2p\gamma)^3}}.$$
The last means that this function has minima for $p = 0$. Due to the fact that this is the degenerate case we can only use $p \approx 0$.

Now we use Berry-Esseen theorem in order to explain how $p$, together with the asymmetry of the distribution{,} influences the accuracy of the estimators. Let us first remind this theorem.

{\bf Berry-Esseen theorem} Let $X_1, X_2, ...$ be i.i.d. r.vs with $EX = a$, $Var\, X = \sigma^2 > 0$ and $E|\frac{X - a}{\sigma}|^3 = r < \infty$ then there exists a positive constant C such that
$$\left|P(\frac{\sqrt{n}(\overline{X}_n - a)}{\sigma} < x) - \Phi(x)\right| \leq C\frac{r}{\sqrt{n}},$$
for all $x \in \mathbb{R}$ and $n \in \mathbb{N}$. Here $\Phi$ is the c.d.f. of the standard normal distribution.

\bigskip

We would like to apply this theorem to the convergence of the sequence $\left\{\frac{1}{k}\sum_{i=1}^{k} U_i^{-\frac{p}{\alpha }}, k = 1, 2, ...\right\}$,
where $\mathbf{U}_1, \mathbf{U}_2, ..., \mathbf{U}_k$ are i.i.d.
uniformly distributed r.v's on $(0, 1)$ and $p \in \mathbb{R}$, to the standard normal distribution. In order to be able to make this we need to impose the restriction the third moment of $U_1^{-\frac{p}{\alpha}}$ to exist, that means that we will consider only $p < \frac{1}{3\gamma} = \frac{\alpha}{3}$ and $p \not = 0$.
{ In that case for $\gamma = 1/\alpha$
$$E \left(\frac{U_1^{-\frac{p}{\alpha}} - EU_1^{-\frac{p}{\alpha}}}{\sqrt{Var\, U_1^{-\frac{p}{\alpha}}}}\right)^3 = \frac{2\sqrt{1 - 2p\gamma}(1 + p\gamma)}{1 - 3p\gamma}.$$

{ The case of $\ln U^{-1/\alpha}$ could be considered again  as a particular case of the above expression for $p = 0$, because it is easy to calculate that
$$E \left(\frac{\ln\,U_1^{-\frac{1}{\alpha}} - E\,\ln\,U_1^{-\frac{1}{\alpha}}}{\sqrt{Var\,\ln\,U_1^{-\frac{1}{\alpha}}}}\right)^3 = 2.$$ }

Let us now calculate the third moment{s} of the absolute values.}

Case 1. $p < 0$. Denote by $a = EU_1^{-\frac{p}{\alpha}}$ and by $\sigma^2 = Var \,EU_1^{-\frac{p}{\alpha}}$.
$$P(U_1^{-\frac{p}{\alpha}} \leq EU_1^{-\frac{p}{\alpha}}) = (EU_1^{-\frac{p}{\alpha}})^{-\frac{\alpha}{p}} = a^{-1/(p\gamma)}$$
$$P(U_1^{-\frac{p}{\alpha}} > EU_1^{-\frac{p}{\alpha}}) = 1 - (EU_1^{-\frac{p}{\alpha}})^{-\frac{\alpha}{p}} = 1 - a^{-1/(p\gamma)}.$$
$$E \left|\frac{U_1^{-\frac{p}{\alpha}} - EU_1^{-\frac{p}{\alpha}}}{\sqrt{Var\, U_1^{-\frac{p}{\alpha}}}}\right|^3 = E \left\{\left(\frac{U_1^{-\frac{p}{\alpha}} - EU_1^{-\frac{p}{\alpha}}}{\sqrt{Var\,U_1^{-\frac{p}{\alpha}}}}\right)^3\left|\right.U_1^{-\frac{p}{\alpha}} > EU_1^{-\frac{p}{\alpha}}\right\} P(U_1^{-\frac{p}{\alpha}} > EU_1^{-\frac{p}{\alpha}}) +$$
$$ + E \left\{\left(\frac{EU_1^{-\frac{p}{\alpha}} - U_1^{-\frac{p}{\alpha}}}{\sqrt{Var\, U_1^{-\frac{p}{\alpha}}}}\right)^3\left| \right. EU_1^{-\frac{p}{\alpha}}> U_1^{-\frac{p}{\alpha}}\right\} P(EU_1^{-\frac{p}{\alpha}} > U_1^{-\frac{p}{\alpha}}) = $$
$$=\frac{1}{|\sigma|^3}\left\{E \left\{\left(U_1^{-\frac{p}{\alpha}} - a\right)^3\left|\right.U_1^{-\frac{p}{\alpha}} > a\right\} (1 - a^{-1/(p\gamma)})+ E \left\{\left(a - U_1^{-\frac{p}{\alpha}}\right)^3\left| \right. a > U_1^{-\frac{p}{\alpha}}\right\} a^{-1/(p\gamma)}\right\} = $$
$$=\frac{\int_0^{a^{1/A}}(a-t^A)^3 dt+\int_{a^{1/A}}^1(t^A-a)^3 dt}{(\frac{A^2}{(1+2A)(1+A)^2})^{3/2}}=$$
 $$=-\frac{2(1 - 2 p\gamma)^{1/2}}{1 - 3 p\gamma}\left[p\gamma + 1 - 6(1-p\gamma)^{1/(p\gamma)-1}\right]:=\phi(p,\gamma)$$

We used that $A:=-p\gamma>0, a:=\frac{1}{1-p\gamma},\ 0<a<1,$
and the fact, that for a random variable with density  $f$ and finite expectation we have  $E(X|a\leq X\leq b)=(\int_a^b f(t)dt)^{-1}\int_a^b tf(t)dt.$

Case 2. $p \in (0 , 1/(3\gamma))$  Using analogous computations we receive the same final formula as in the previous case.

{Case 3. $p = 0$, the Hill estimator.
$$E \left|\frac{\ln\, U_1^{-\frac{1}{\alpha}} - E\,\ln\,U_1^{-\frac{1}{\alpha}}}{\sqrt{Var\,\ln\, U_1^{-\frac{1p}{\alpha}}}}\right|^3 =
\frac{12}{e} - 2=\lim_{p\to 0} \phi(p,\gamma)\approx 2.4146.$$
Note that this value does not depend on $\alpha$.}

\begin{figure}
\begin{center}
\includegraphics[scale=.33,angle =-90 ]{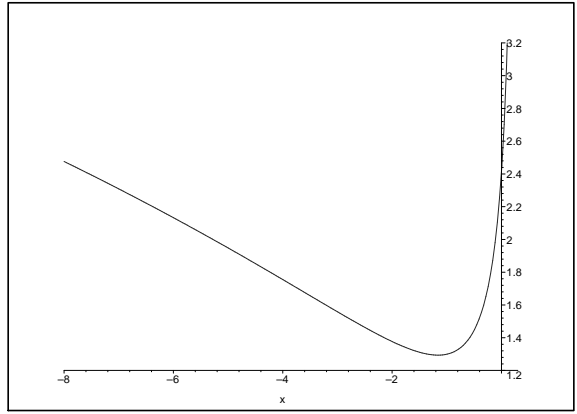}
\caption{Plot of $ E\left|\frac{U_1^{-x} - EU_1^{-x}}{\sqrt{Var\, U_1^{-x}}}\right|^3$}
\label{fig:Skew}
\end{center}
\end{figure}

The function $\phi(p,\gamma)$ is plotted at range $p\gamma\in (-8, 0.3)$ at the Figure \ref{fig:Skew}.
So, we can conclude that if we minimize the variance of the estimators we should chose
\begin{equation}\label{HillOpt}
p_H = 0,
\end{equation} and this corresponds to the Hill-estimator.
If we would like to chose $p$ in such a way in order  to have smallest distance between the distributions and determine this closeness by Berry-Esseen theorem for  fixed $k$, then we receive optimal $p$ by
\begin{equation}\label{SkewOpt}
p_{opt}(\alpha)\approx -1.221\alpha.
\end{equation}
Also from the Figure \ref{fig:Skew} we see that function $\phi(p,\gamma)$ is convex and that
we obtain the interval $-7.64<p\gamma<0$ for such $p$ parameters which solve  $\phi(p,\gamma)<\phi(0,\gamma),$ where generalized Hill estimator is better that the Hill in the sense of Berry-Esseen.

 \cite{PaulauskasVaiciulis} works under Mason condition $k(n) = o(n)$.
Their estimator $\hat \gamma_n^{(1)}(k,r)$ coincides with Harmonic mean estimator of \cite{BeranSSt}.
The relationship is $ 1-p=\beta=1-r.  $ In case of $k_n = o(n)$ and $\overline{F} \in 2RV_{-\alpha, \rho}$, $\rho < 0$
\cite{PaulauskasVaiciulis} obtained for $\hat \gamma_n^{(1)}(k,r)$
 an explicit (and rather simple) expressions
for the optimal value of the parameter $p$. Their formula is
\begin{equation}\label{Paulaskas}
p^* = \frac{2-\rho\gamma - \sqrt{(2-\rho\gamma)^2 - 2}}{2\gamma}.
\end{equation}

It is easy to check that using $p^*$ from formula $(\ref{Paulaskas})$ is not giving an optimal value of $p$ for examples 1-3 neither in the sense of Berry Esseen approximation (see \ref{SkewOpt}), nor the optimal variance for Hill-estimator (\ref{HillOpt}),  since it overestimates $p$. Indeed, we have   \begin{itemize}
\item for Example 1, for $HW(1,-1)$ we have $$p^*=\frac32-\frac{\sqrt{7}}{2}\approx 0.177>p_H>p_{opt}(1)=-1.221.$$

\item for Example 1, for $HW(2,-1)$ we have $$p^*=\frac52-\frac{\sqrt{17}}{2}\approx 0.438>p_H>p_{opt}(2)=-2.442.$$

\item for Example 1, for $HW(2,-1)$ we have $$p^*=2-\frac{\sqrt{14}}{2}\approx 0.129>p_H>p_{opt}(1)=-1.221.$$

\item for Example 2, we have $\rho=0,\alpha=1$ and thus we obtain  $$p^*=1-\frac{\sqrt{2}}{2}\approx 0.292>p_H>p_{opt}(1)=-1.221.$$

\end{itemize}

Actually, it can be easily checked  for $(\ref{Paulaskas})$ that $p^*\geq 0$ for all $\alpha>0,\rho<0.$

\section{Empirical investigation}
Let us assume that the observed r.v. $X$ has  continuous theoretical c.d.f. $F$ which is in the max-domain of attraction of some extreme value distribution with parameter $\gamma \in \mathbb{R}$. From the Theorem 7 in \cite{Pickands} we see that this d.f. has generalized Pareto upper tail with parameter $\gamma$. Due to the laws of the zero and once we can say that if we have independent observations of some random variable $X$ and if we have enough data then the Pareto tail behavior will always appear in the data.

 \bigskip

In this part we show that the Pareto tail behavior could be observed in usual real data set with not too much observations and we do not need to check for the second order regular variation condition in order to use the asymptotic normality and to obtain confidence intervals of the index of regular variation. The data that we use here are taken from recent study of snow extremes in Slovakia (see \cite{Snow}).
 The observed random variable $k$ explains {the ratio of the snow load to the characteristic snow load.} In the spirit of the reproducible research and because of the sample size is only $n = 41$ we present also the data set.

 \begin{verbatim}
  2.03, 2, 2, 1.96, 1.83, 1.83, 1.80, 1.78, 1.75, 1.75, 1.75, 1.75,
  1.73, 1.71, 1.71, 1.67, 1.67, 1.66, 1.65, 1.65, 1.65, 1.65, 1.64,
  1.63, 1.61, 1.6,  1.60, 1.60, 1.59, 1.58, 1.56, 1.56, 1.55, 1.53,
  1.53, 1.51, 1.5,  1.49, 1.49, 1.49, 1.49
 \end{verbatim}

 The mean excess function of the data is given on Figure \ref{fig:MEanE}. It shows that appropriate choice of the threshold above which we can consider Pareto behavior of the data is $u = 1.65$. We observe that $18$ observations exceed this threshold.

\bigskip

\begin{figure}[h]
\begin{center}
\begin{minipage}[t]{0.45\linewidth}
\caption{Mean excess plot}
\includegraphics[scale=.4]{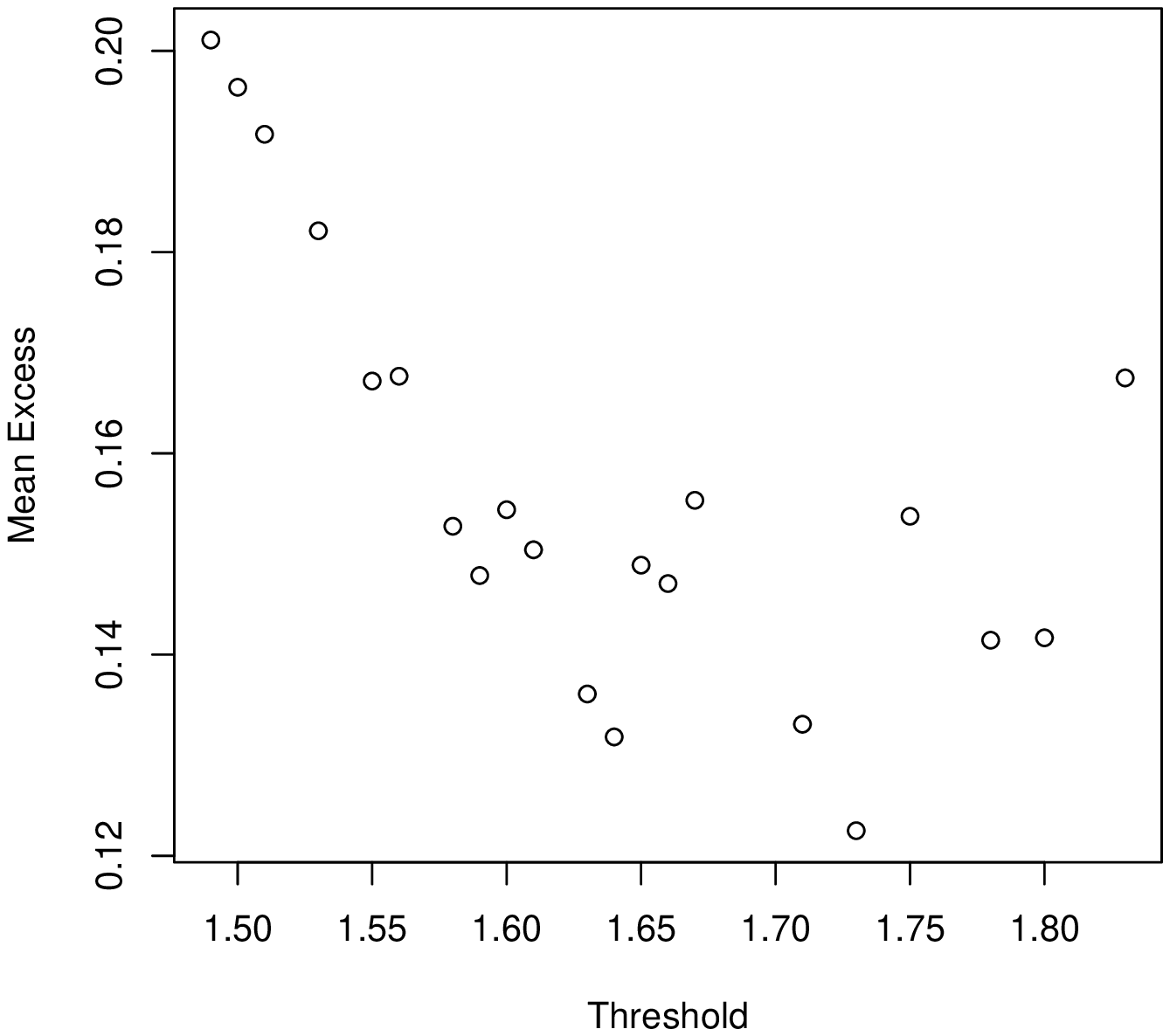}
    \label{fig:MEanE}\end{minipage}
\begin{minipage}[t]{0.45\linewidth}
\caption{Generalized Hill and Hill plots}
\includegraphics[scale=.4]{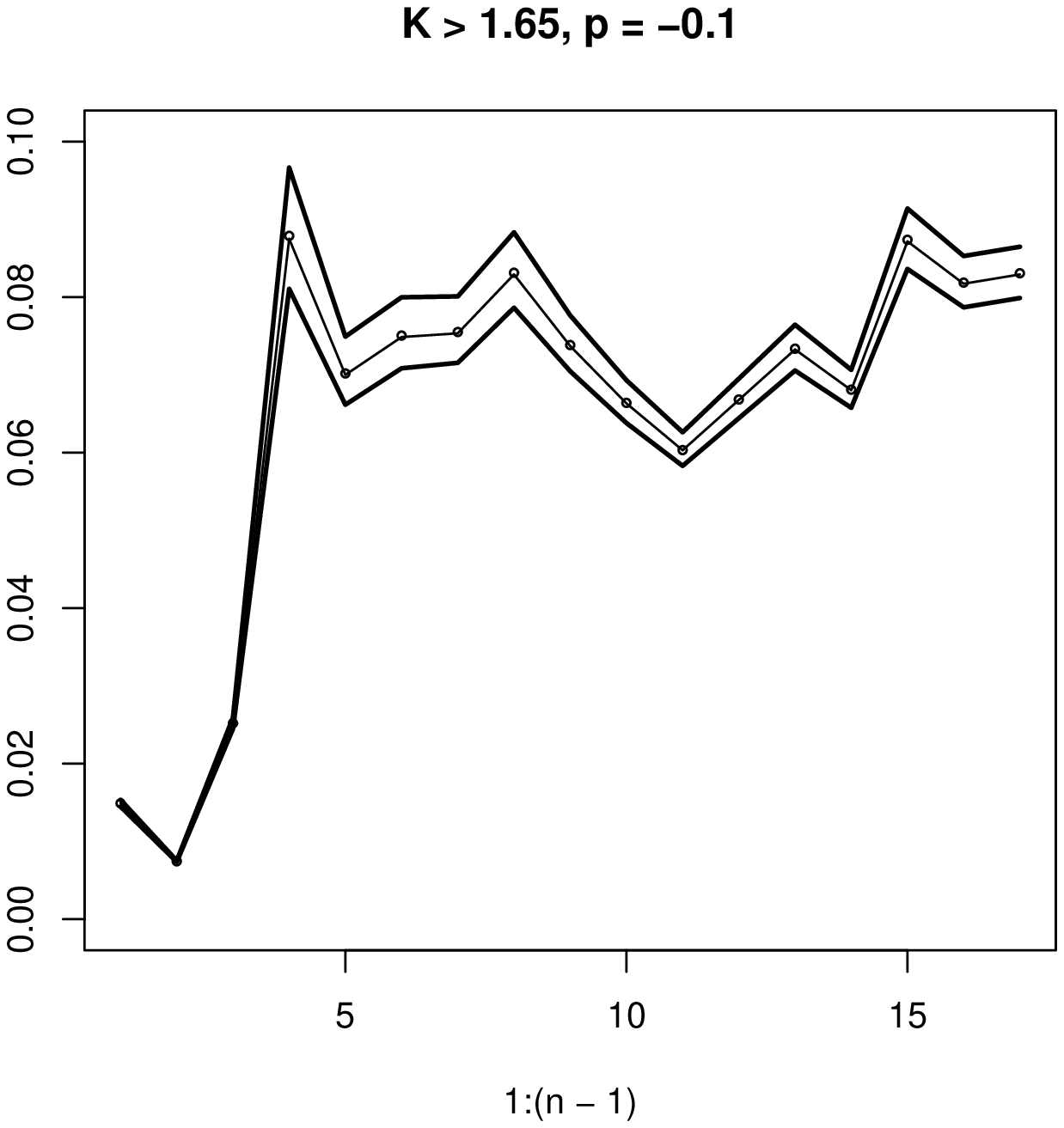}
    \label{fig:GHP}
\end{minipage}
\end{center}
\end{figure}

\bigskip

The Generalized Hill and Hill plots for $p = -0.1$, together with the Hill estimator and its confidence intervals, based on the normal approximation (see formula (\ref{conf_int})), are given on Figure \ref{fig:GHP}.  The Hill estimator for $k = 17$  delivered estimator $\hat\gamma=0.0829$ for value of $\gamma$ and 0.95 confidence interval is $(0.0799, 0.0865)$. Its Generalized Hill estimator for $p = -0.1$ is $0.0831$. Now we can use the peaks over threshold technique for the estimation of the high quantiles and obtain that for $x > 1.65$
$$\hat{P}(K > x) = \left(\frac{x}{1.65}\right)^{-1/0.0829}\frac{18}{41}$$
The last means e.g. that the level 2.5 will be exceeded approximately 2.4 times in 1000 years (see \cite{Pickands}).

{\section{Discussion and conclusions}

In this paper we illustrated a flexible approach for extreme value modelling.
In particular, we have proven asymptotic normality without 2nd order regularly varying condition, suitable
for a small samples or complicated practical examples. We have also illustrated theoretically that 2nd order regularly varying
condition is not necessary for asymptotic normality.
The alternative requirements on design for such samples are needed. Namely, we do not expect discovery of Pareto tail in arbitrary small amount of data e.g. less than 30 unless these data come from exact Pareto distribution. In the last case usually small amount of data, more than 30, may be enough because in such case  the normal approximation works relatively well.
We cannot apply generalized Hill estimators when we have no RV tails.
However, early or soon any subset of data { that come from independent observations of  r.v. with c.d.f. with regularly varying tail} will show its Pareto tail behavior if we have enough data.
From practical point of view, enough data means its mean excess function to become increasing from some point further on.
In such case in order to estimate Pareto tail we take only the biggest observations.

\bibliographystyle{imsart-nameyear}

\end{document}